\documentclass[11pt]{article}
\usepackage{amssymb}
\usepackage{amsmath}
\usepackage{amscd}
\usepackage{latexsym}
\usepackage{graphicx}

\catcode`@=11
\renewcommand{\section}{\@startsection{section}{1}{0pt}{\medskipamount}
{\medskipamount}{\large\bf}}
\numberwithin{equation}{section}
\catcode`@=12

\def\m{\mu}
\def\n{\nu}
\newcommand{\Z}{\mathbb Z}
\newcommand{\C}{\mathbb C}

\newcommand{\Hb}{{\mathbb H}}
\newcommand{\R}{\mathbb R}
\newcommand{\Acal}{{\cal A}}
\newcommand{\Fcal}{{\cal F}}
\newcommand{\Ecal}{{\cal E}}
\newcommand{\rt}{\textrm{t}}
\newcommand{\rr}{\textrm{r}}

\newcommand{\Fct}{{\widetilde{\cal F}}}
\newcommand{\Act}{{\widetilde{\cal A}}}
\newcommand{\Et}{{\widetilde{E}}}
\newcommand{\te}{{\tilde{e}}}
\newcommand{\Bt}{{\widetilde{B}}}
\newcommand{\al}{\alpha}
\newcommand{\be}{\beta}
\newcommand{\ga}{\gamma}
\newcommand{\de}{\delta}

\newcommand{\ve}{\varepsilon}
\newcommand{\vp}{\varphi}
\renewcommand{\th}{\theta}
\newcommand{\la}{\lambda}
\newcommand{\ph}{\phi}
\newcommand{\ch}{\chi}
\newcommand{\om}{\omega}
\newcommand{\we}{{\wedge}}
\def\im{\mathrm{i}}
\def\ep{\mathrm{e}}
\def\N2{$N{=}2$}
\def\pa{\mbox{$\partial$}}
\def\diff{\mathrm{d}}
\def\tr{\mathrm{tr}}
\def\sfrac#1#2{{\textstyle\frac{#1}{#2}}}
\def\>{\rangle}
\def\<{\langle}
\def\+{\dagger}
\def\={\ =\ }
\def\und{\quad\textrm{and}\quad}
\def\with{\quad\textrm{with}\quad}
\def\for{\quad\textrm{for}\quad}

\def\Id{\mathrm{Id}}

\newcommand{\beq}{\begin{equation}}
\newcommand{\eeq}{\end{equation}}
\newcommand{\bea}{\begin{eqnarray}}
\newcommand{\eea}{\end{eqnarray}}

\begin{document}

\begin{titlepage}
\setcounter{page}{0}


\begin{center}
{\LARGE{\bf 
Finite-action solutions of Yang--Mills equations \\[12pt]
on de Sitter dS$_4$ and anti-de Sitter AdS$_4$ spaces}}

\vspace{10mm}

{\Large Tatiana A. Ivanova${}^*$, \ Olaf Lechtenfeld${}^{\+}$ \ and \  Alexander D. Popov${}^\+$
}\\[10mm]
\noindent ${}^*${\em
Bogoliubov Laboratory of Theoretical Physics, JINR\\
141980 Dubna, Moscow Region, Russia
}\\
{Email: ita@theor.jinr.ru
}\\[5mm]
\noindent ${}^\+${\em Institut f\"ur Theoretische Physik \,{\rm and} 
Riemann Center for Geometry and Physics\\
Leibniz Universit\"at Hannover \\
Appelstra\ss{}e 2, 30167 Hannover, Germany
}\\
{Email: olaf.lechtenfeld@itp.uni-hannover.de, alexander.popov@itp.uni-hannover.de}

\vspace{15mm}

\begin{abstract}
\noindent We consider pure SU(2) Yang--Mills theory on four-dimensional de Sitter dS$_4$ and anti-de Sitter AdS$_4$ spaces and
construct various solutions to the Yang--Mills equations. On de Sitter space we reduce the Yang--Mills equations via
an SU(2)-equivariant ansatz to Newtonian mechanics of a particle moving in $\R^3$  under the influence of a quartic potential.
Then we describe magnetic and electric-magnetic solutions, both Abelian and non-Abelian, all having finite energy and finite
action. A similar reduction on anti-de Sitter space also yields Yang--Mills solutions with finite energy and action. 
We propose a lower bound for the action on both backgrounds. Employing another metric on AdS$_4$, the SU(2) Yang--Mills equations are 
reduced to an analytic continuation of the above particle mechanics from $\R^3$ to $\R^{2,1}$. We discuss analytical solutions 
to these equations, which produce infinite-action configurations. After a Euclidean continuation of dS$_4$ and AdS$_4$ we also 
present self-dual (instanton-type) Yang--Mills solutions on these backgrounds.
\end{abstract}

\vspace{12mm}


\end{center}
\end{titlepage}

\section{Introduction}

\noindent 
Magnetic monopoles~\cite{Dirac} and vortices~\cite{JT} are playing an important role in the nonperturbative physics of
$3{+}1$ dimensional Yang--Mills--Higgs theory~\cite{Raj, MS, Wein}. However, in pure gauge theory without any scalar fields
there are no vortices or non-Abelian monopoles on Minkowski space~$\R^{3,1}$. Yet, our universe appears to be asymptotically 
de Sitter (not Minkowski) at very early and very late times. This provides strong motivation for searching finite-action
solutions in pure Yang--Mills theory on de Sitter space dS$_4$. Finding Yang--Mills solutions on anti-de~Sitter space AdS$_4$
is also reasonable from the viewpoint of string-theory applications and from the AdS/CFT perspective. The construction of such
solutions, both Abelian and non-Abelian, is the goal of our paper.\footnote{ 
We consider the spacetime background as non-dynamical, i.e.~we ignore the backreaction on it. The coupled system is governed 
by the Einstein--Yang--Mills equations (for numerical solutions, see e.g.~the review~\cite{KK} and references therein). 
However, in such a more general setup it is practically impossible to obtain analytic non-Abelian solutions.} 
Some steps in this direction have been made in~\cite{IvLePo}.

In this paper, we present a construction of smooth Abelian and non-Abelian solutions with both finite energy and action in
pure Yang--Mills theory on de Sitter space dS$_4$ and anti-de Sitter space AdS$_4$. Other types of Yang--Mills solutions on  
anti-de Sitter space, also described in this paper, have infinite energy and action. We also write down instantons and
quasi-instantons in de Sitter dS$_4$ and anti-de Sitter AdS$_4$ spaces.
We postpone the issue of boundary conditions and study classical solutions for any kind of boundary condition.

The paper is organized as follows. Section~2 provides a description of de Sitter space, which is used in Section~3 to
explicitly construct Yang--Mills solutions on dS$_4$ and to compute their energy and action. Instantons on the Euclideanized
background are the topic of Section~4. The story is repeated for anti-de Sitter space AdS$_4$ in Sections 5, 6 and~7.
We conclude in Section~8. Four appendices list the various metrics used in the paper, detail metrics on the spatial slices
$S^3$ and AdS$_3$, and present explicit expressions for our Yang--Mills solutions in the fundamental and in the adjoint
SU(2) representation on dS$_4$ in various coordinates.

\bigskip

\section{Description of de Sitter space dS$_4$}

\noindent {\bf Closed-slicing coordinates.}
Four-dimensional de Sitter space can be embedded into five-di\-men\-sion\-al
Minkowski space $\R^{4,1}$ as the one-sheeted hyperboloid
\begin{equation}\label{2.1}
\de_{ij}y^iy^j - (y^5)^2 \= R^2 \qquad\textrm{where}\quad i,j=1,\ldots,4\ .
\end{equation}
Topologically, de Sitter space dS$_4$ is $\R\times S^3$, and one
can introduce global coordinates $(\tau , \chi , \th , \phi)$
adapted to this topology by setting (see e.g. \cite{HE})
\begin{equation}\label{2.2}
y^i= R\,\om^i\cosh\tau\ ,\quad y^5=R\,\sinh\tau
\qquad\textrm{with}\quad\tau\in\R
\und \de_{ij}\,\om^i\om^j=1
\end{equation}
for $\om^i=\om^i(\chi,\th,\phi)$ embedding $S^3$ into $\R^{4,0}$.
A dimensionful time coordinate may be introduced as $\tilde\tau=R\,\tau$.
The flat metric on $\R^{4,1}$ induces a metric on dS$_4$,
\beq\label{2.4}
\diff s^2 \= R^2\,\bigl( -\diff\tau^2 + \cosh^2\!\tau\,\diff\Omega_3^2 \bigr)
\eeq
with $\diff\Omega_3^2$ being the metric on the unit sphere $S^3\cong\textrm{SU}(2)$.

On this unit $S^3$ we introduce an orthonormal basis $\{e^a\}\, ,\ a=1,2,3,$ of left-invariant one-forms satisfying
\begin{equation}\label{2.5}
\diff e^a + \ve^a_{bc}\,e^b\wedge e^c\=0\ .
\end{equation}
For an embedding $\{\omega^i\}$, the one-forms $\{e^a\}$ can be constructed via
\beq\label{2.6}
e^a \= -\eta^a_{ij}\,\omega^i\diff\omega^j\ ,\qquad\textrm{with}\quad
\eta^a_{ij} \= \begin{cases}
\ve^{a}_{ij} & \textrm{for} \quad i,j=1,2,3 \\[-2pt]
{+}\de^{a}_{i} & \textrm{for} \quad j=4 \\[-2pt]
{-}\de^{a}_{j} & \textrm{for} \quad i=4 \\[-2pt]
0 & \textrm{for} \quad i=j=4 \end{cases}
\eeq
denoting the self-dual 't Hooft symbols.
The metric on $S^3$ is then obtained as
\begin{equation}\label{2.7}
\diff\Omega_3^2 \= \bigl(e^1\bigr)^2 + \bigl(e^2\bigr)^2 + \bigl(e^3\bigr)^2\ .
\end{equation}
In Appendix~B we explicitly present two prominent such embeddings
and the corresponding one-forms and metric.

\smallskip

\noindent {\bf Conformal coordinates.}
One can rewrite the metric (\ref{2.4}) on dS$_4$ in
conformal coordinates $(t,\chi , \th , \phi )$ by the time reparametrization~\cite{HE}
\begin{equation}\label{2.14}
t\=\arctan (\sinh\tau)\=2\arctan(\tanh\sfrac{\tau}{2})
\qquad\Longleftrightarrow\qquad
\frac{\diff\tau}{\diff t} \= \cosh\tau \= \frac{1}{\cos t}\ ,
\end{equation}
in which $\tau\in (-\infty, \infty )$ corresponds to $t\in (-\sfrac{\pi}{2}, \sfrac{\pi}{2})$.
The metric (\ref{2.4}) in these coordinates reads
\begin{equation}\label{2.15}
\diff s^2 \= \frac{R^2}{\cos^2\!t}\,\bigl(- \diff t^2 + \de_{ab}e^ae^b \bigr)
\= \frac{R^2}{\cos^2\!t}\,\diff s^2_{\textrm{cyl}}  \ ,
\end{equation}
where
\begin{equation}\label{2.16}
\diff s^2_{\textrm{cyl}} \= - \diff t^2 + \de_{ab}e^ae^b
\end{equation}
is the standard metric on the Lorentzian cylinder $\R\times S^3$.
Hence, four-dimensional de Sitter space is conformally equivalent to the finite cylinder
${\cal I}\times S^3$ with the metric (\ref{2.16}),
where ${\cal I}$ is the interval $(-\frac{\pi}{2}, \frac{\pi}{2})$ parametrized by~$t$.

\smallskip

\noindent 
{\bf Static coordinates.} 
The sphere $S^3$ can be glued from a pair $(S^3_+,S^3_-)$ of 3-balls and a 2-sphere~$S^2$,
\begin{equation}\label{2.10}
S^3\= S^3_+\cup S^2\cup S^3_- \ ,
\end{equation}
where $S^3_+$ is an `upper hemisphere',  $S^3_-$ is the `lower hemisphere', and the gluing surface\footnote{
See Appendix~B for more details. In particular, for the metric (\ref{A.2}) we may take the angle 
$0\le\chi<\frac{\pi}{2}$ for $S^3_+$, $\frac{\pi}{2}<\chi\le\pi$ for $S^3_-$ and $\chi=\frac{\pi}{2}$ for the equatorial $S^2$.} 
is the equatorial 2-sphere $S^2$. On any `half' $\R\times S^3_\pm \cong\R^4$ of dS$_4$ one may introduce static coordinates 
$(\sigma,\rho,\th,\ph)$ by taking 
\beq\label{2.17} 
y^a = R\,\rho\,\la^a\, ,\ y^4 = R\,\sqrt{1{-}\rho^2}\,\cosh\sigma\, , \ y^5 = R\,\sqrt{1{-}\rho^2}\,\sinh\sigma 
\with \sigma\in\R\, ,\ \rho\in[0,1) 
\eeq 
and $\de_{ab}\la^a\la^b=1$ for 
\beq\label{2.18} 
\la^1=\sin\th\,\sin\ph\ ,\quad \la^2=\sin\th\,\cos\ph\ ,\quad \la^3=\cos\th\ . 
\eeq 
In this case, the induced metric on dS$_4$ comes out as 
\beq\label{2.19} 
\diff s^2 \= R^2\,\bigl( -(1{-}\rho^2)\diff\sigma^2 + \sfrac{\diff\rho^2}{1{-}\rho^2} + \rho^2\diff\Omega_2^2 \bigr) 
\quad\with \diff\Omega_2^2 \= \diff\th^2+\sin^2\!\th\,\diff\ph^2\ . 
\eeq 
Dimensionful time and radial coordinates are $\rt=R\,\sigma$ and $\rr=R\,\rho$.
Obviously, the metric (\ref{2.19}) has a cosmological event horizon at $\rho{=}1$ (or $\rr{=}R$) and, therefore, static
coordinates cover only half of dS$_4$. In its range it is convenient to introduce a coordinate $\alpha$ instead of $\rho$ via
\beq\label{2.20} 
\rho=\sin\al \qquad\Rightarrow\qquad \sqrt{1{-}\rho^2}=\cos\al\ , 
\eeq 
that will be used in later calculations.

\bigskip

\section{Yang--Mills configurations on dS$_4$}

\noindent
In Minkowski space $\R^{3,1}$ smooth vortex or monopole solutions of gauge
theory can be constructed only in the presence of Higgs fields. Both types of
solutions have finite energy per length or energy, defined via integrals over
$\R^{2,0}\subset\R^{3,1}$ or $\R^{3,0}\subset\R^{3,1}$, respectively.
There are no finite-energy solutions
of such kind in pure Yang--Mills theory in Minkowski space. Here, we will show that
finite-energy solutions in gauge theory without scalar fields do exist on de Sitter
space dS$_4$. Furthermore, they also have finite action, contrary to monopoles
or vortices in $\R^{3,1}$.

\smallskip

\noindent {\bf Conformal invariance.}
Since in four dimensions the Yang--Mills equations are conformally
invariant, their solutions on de Sitter space can be obtained by solving
the equations on ${\cal I}\times S^3$ with the cylindrical metric (\ref{2.16}).
Therefore, we will consider a rank-$N$ Hermitian vector bundle over the cylinder
${\cal I}\times S^3$ with a gauge potential $\Acal$ and the gauge field
$\Fcal=\diff\Acal + \Acal\wedge\Acal$ taking values in the Lie algebra $su(N)$.
 The conformal boundary of dS$_4$ consists of the two 3-spheres at
$t=\pm\frac{\pi}{2}$ or, equivalently, at $\tau=\pm\infty$.
On a manifold $M$ with boundary $\pa M$, gauge transformations
are naturally restricted to tend to the identity when approaching $\pa M$ (see e.g.~\cite{Don}).
This corresponds to a framing of the gauge bundle over the boundary.
For our case, it means allowing only gauge-group elements $g(\cdot)$ subject to
\begin{equation}
g(\pa M)\=\textrm{Id} \qquad\textrm{on}\quad
\pa M=S^3_{t={+}\frac{\pi}{2}}\cup S^3_{t={-}\frac{\pi}{2}}\ .
\end{equation}

\smallskip

\noindent {\bf Reduction to matrix equations.}
In order to obtain explicit solutions
we use the SU(2)-equivariant ansatz (cf.~\cite{IL, ILPR, BILL})
\begin{equation}\label{3.1}
\Acal = X_a(t)\,e^a
\end{equation}
for the $su(N)$-valued gauge potential $\Acal$ in the temporal gauge
$\Acal_0\equiv\Acal_t=0=\Acal_\tau$. Here, $X_a(t)$ are three
$su(N)$-valued functions depending only on $t\in{\cal I}$, and $e^a$
are the basis one-forms on $S^3$ satisfying (\ref{2.5}).
The corresponding gauge field reads
\begin{equation}\label{3.2}
\Fcal \= \Fcal_{0a}\,e^0\wedge e^a + \sfrac12 \Fcal_{bc}\,e^b\wedge e^c
\= \dot{X}_a\,e^0\wedge e^a + \sfrac12\bigl(-2\ve^a_{bc}X_a + [X_b, X_c]\bigr)e^b\wedge e^c\ ,
\end{equation}
where $\dot{X}_a:=\diff{X}_a/\diff t$ and $e^0:=\diff t$.
It is not difficult to show (see e.g.~\cite{BILL}) that the Yang--Mills equations on ${\cal I}\times S^3$
after substituting (\ref{3.1}) and (\ref{3.2}) reduce to the ordinary matrix differential equations
\begin{equation}\label{3.3}
\ddot X_a \= -4 X_a+ 3 \ve_{abc}\, [X_b, X_c] - \bigl [X_b, [X_a,X_b]\bigr]
\qquad\und\qquad [X_a,\dot{X}_a]\= 0\ .
\end{equation}

\smallskip

\noindent {\bf Reduction to a Newtonian particle on $\R^3$.}
It is very natural to restrict the matrices $X_a$ to an $su(2)$ subalgebra. To this end, we
embed the spin-$j$ representation of SU(2) into the fundamental of SU($N$) with $N=2j{+}1$.
The three SU(2) generators $I_a$ then obey
\begin{equation}\label{3.4}
[I_b, I_c] \= 2\,\ve^a_{bc}I_a \quad\und\quad \tr(I_aI_b)\=-4\,C(j)\,\de_{ab} \quad\for
C(j)=\sfrac13\,j(j{+}1)(2j{+}1)\ ,
\end{equation}
where $C(j)$ is the second-order Dynkin index of the spin-$j$ representation.
The simplest choice for $X_a$ then is\,\footnote{
This resolves the second equation in (\ref{3.3}), the first-order Gau\ss-law constraint.
For a more general form of $X_a$, related with $A_k$-quivers, see e.g.~\cite{ILPS, LPS}.}
\begin{equation}\label{3.5}
X_1=\Psi_1 I_1\ ,\quad X_2=\Psi_2 I_2\und X_3=\Psi_3 I_3\ ,
\end{equation}
where $\Psi_a$ are real functions of $t\in\cal I$.

Due to the equivalence of Yang--Mills theory on dS$_4$ with metric (\ref{2.15})
to the theory on ${\cal I}\times S^3$ with metric (\ref{2.16}),
we obtain the Lagrangian density
\beq\label{3.6}
\begin{aligned}
{\cal L} &\= \sfrac18\,\tr \Fcal_{\mu\nu}\Fcal^{\mu\nu}\=
-\sfrac14\,\tr\Fcal_{0a}\Fcal_{0a}+\sfrac18\,\tr\Fcal_{ab}\Fcal_{ab} \\[4pt]
&\= 4\,C(j)\bigl\{\sfrac14\dot{\Psi}_a\dot{\Psi}_a - (\Psi_1-\Psi_2\Psi_3)^2-
(\Psi_2-\Psi_3\Psi_1)^2 - (\Psi_3-\Psi_1\Psi_2)^2\bigr\}\ .
\end{aligned}
\eeq
The $S^3$ has disappeared, and we are left with at a Lagrangian on~$\cal I$.
Interpreting the real functions $\Psi_a(t)$ as coordinates of a particle on~$\R^3$,
this Lagrangian describes its Newtonian dynamics in a finite time interval,
with kinetic energy~$T$ and quartic potential energy~$V$,\footnote{
Interestingly, $V=\sfrac12 \pa_a U\pa_a U$ with a superpotential
$U=\Psi_1^2{+}\Psi_2^2{+}\Psi_3^2-2\Psi_1\Psi_2\Psi_3$,
but for Minkowski time this does not yield a gradient flow.}
\beq\label{3.7}
T\=\sfrac12\,\dot{\Psi}_a\dot{\Psi}_a \quad\und\quad
V\=2\,\bigl\{(\Psi_1-\Psi_2\Psi_3)^2+(\Psi_2-\Psi_3\Psi_1)^2+(\Psi_3-\Psi_1\Psi_2)^2\bigr\}\ .
\eeq
\begin{figure}[ht!]
\centering
\includegraphics[scale=1.0]{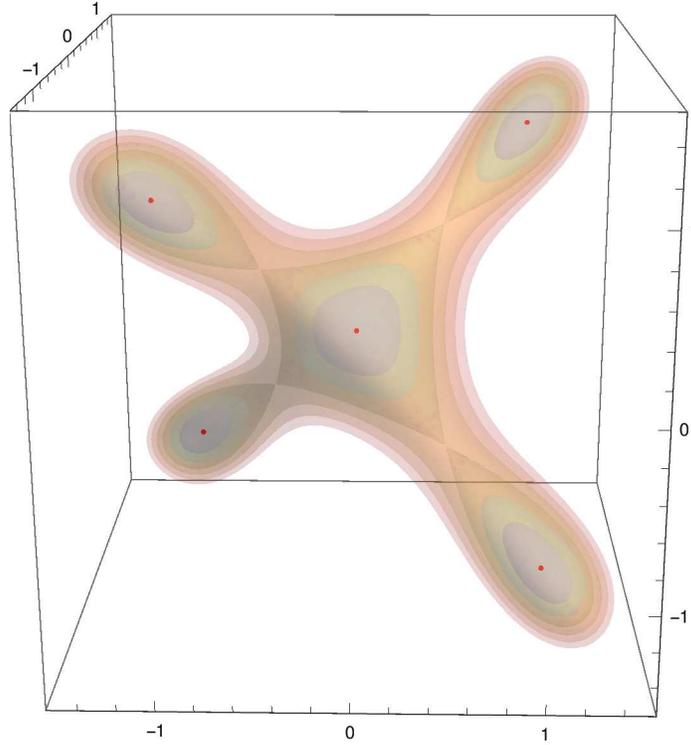}
\caption{Contours of the Newtonian potential $V$ in (\ref{3.7}).}
\label{fig1}
\end{figure}
The critical points $(\hat\Psi_1,\hat\Psi_2,\hat\Psi_3)$ of this potential are
\beq \label{critpts}
(0,0,0) \ =\textrm{minimum}\quad,\qquad
(\pm\sfrac12,\pm\sfrac12,\pm\sfrac12) \ =\textrm{saddle}\quad,\qquad
(\pm1,\pm1,\pm1) \ =\textrm{minima}\ ,
\eeq
with
\beq
V(\textrm{minima}) = 0 \qquad\und\qquad  V(\textrm{saddle}) = \sfrac38\ ,
\eeq
where the number of minus signs in each triple must be even.
The central minimum is isotropic with oscillation frequency $\om=2$.
The other four minima support eigenoscillations with frequencies $\om_\parallel=2$
and $\om_\perp=4$ with respect to the radial direction.

The equations of motion can be obtained either by substituting (\ref{3.5}) into (\ref{3.3})
or from (\ref{3.6}) as the Euler--Lagrange equations,
\beq\label{3.8}
\begin{aligned}
\sfrac14\ddot{\Psi}_1&\= -\Psi_1 + 3\,\Psi_2\Psi_3 - \Psi_1(\Psi_2^2 + \Psi_3^2)\ ,\\
\sfrac14\ddot{\Psi}_2&\= -\Psi_2 + 3\,\Psi_1\Psi_3 - \Psi_2(\Psi_1^2 + \Psi_3^2)\ ,\\
\sfrac14\ddot{\Psi}_3&\= -\Psi_3 + 3\,\Psi_1\Psi_2 - \Psi_3(\Psi_2^2 + \Psi_3^2)\ .
\end{aligned}
\eeq

These equations are still difficult to solve. However, as can be seen from the contour plot of the potential in Fig.~1, 
the system enjoys tetrahedral symmetry. The permutation group $S_4$ acts on the triple $(\Psi_1,\Psi_2,\Psi_3)\in\R^3$ by 
permuting the entries and by changing the sign of an even number of entries. One may hope to find analytic solutions for trajectories 
fixed under part of this symmetry. The maximal subgroups of $S_4$ are $A_4$ (of order 2), $D_8$ (of order 3) and $S_3$ (of order 4).
While $A_4$ leaves only the origin invariant, $D_8$ keeps fixed a coordinate axis (up to sign), and $S_3$ leaves invariant the
direction to a noncentral potential minimum. Therefore, we look at two special cases. 
In the $D_8$ case, we pick the $\Psi_3$-axis and consider
\beq\label{3.10} 
\Psi_1 = \Psi_2 = 0\und \Psi_3 =\xi\ , \eeq 
where $\xi(t)$ is some real-valued function of $t\in (-\sfrac{\pi}{2}, \sfrac{\pi}{2})$. 
With this simplification, we get
\beq\label{3.14}
T_{\xi}=\sfrac12\,\dot{\xi}^2\ ,\quad
V_{\xi}=2\,{\xi}^2 \quad\und\quad \ddot{\xi} \= -4\,\xi \ , 
\eeq
showing that in this direction in parameter space the harmonic approximation is exact.
Here, the non-stable $S_4$ transformations act by permuting the coordinate axes, 
but two equivalent choices give the same equations.
In the $S_3$ case, we choose the direction $(1,1,1)$ and put 
\beq\label{3.9} 
\Psi_1 = \Psi_2 = \Psi_3 = \sfrac12 (1+\psi)\ , 
\eeq 
where $\psi(t)$ is some other real-valued function of $t\in (-\sfrac{\pi}{2}, \sfrac{\pi}{2})$. 
This ansatz leads to the simplifications
\beq\label{3.12}
T_{\psi}=\sfrac12\,\dot{\psi}^2\ ,\quad V_{\psi}=\sfrac12\,(1{-}{\psi}^2)^2 \quad\und\quad
\ddot{\psi}\=2\,\psi\,(1{-}\psi^2)\ . \eeq
The remaining $S_4$ transformations flip the sign of two coordinates, which
generates three other but equivalent configurations, yielding the same equations. 
Other directions fixed under some subgroup of~$S_4$ do not give rise to elementary solutions.

\smallskip

\noindent {\bf Solutions.}
The general solution of the linear equation~(\ref{3.14}) is
\begin{equation} \label{3.18}
\xi(t) \= -\sfrac{\ga}{2}\,\cos 2(t{-}t_0) \ ,
\end{equation}
where $\ga$ and $t_0$ are arbitrary real parameters.
Since only one $su(2)$ generator is excited, it leads to an Abelian field configuration.
We note that the normalization of the linear solution $\xi(t)$ is arbitrary.

For the non-Abelian ansatz (\ref{3.9}) the simplest solutions of~(\ref{3.12}) 
are constant at the critical points of~$V_{\psi}$, i.e.
\begin{equation} \label{3.15}
\psi(t) = \pm1 \ \textrm{(minima, $V_{\psi}{=}0$)} \und
\psi(t) = 0 \ \textrm{(local maximum, $V_{\psi}{=}\sfrac12$)}\ .
\end{equation}
A prominent nontrivial solution of (\ref{3.12}) is the bounce,
\begin{equation} \label{3.16}
\psi(t)\= \sqrt{2}\,\textrm{sech}\bigl(\sqrt{2}(t{-}t_0)\bigr)
\= \frac{\sqrt{2}}{\cosh\bigl(\sqrt{2}(t{-}t_0)\bigr)}\ ,
\end{equation}
which describes the motion from the local maximum $(\psi{=}0)$ at $t{=}{-}\infty$
on the cylinder $\R\times S^3$ to the turning point $(\psi{=}\sqrt{2})$ at $t{=}t_0$
and back to $(\psi{=}0)$ at $t{=}\infty$.
Flipping the sign of $\psi(t)$ produces the anti-bounce, which explores the other half
of the double-well potential. In addition, there is a continuum of periodic solutions
oscillating either about $\psi=\pm 1$ or exploring both wells of the double-well
potential $V_{\psi}$, which are given by Jacobi elliptic functions. Usually, the moduli parameter
$t_0$ is trivial because of time translation invariance in~(\ref{3.12}).
However, since for de Sitter space according to~(\ref{2.14}) we consider the solutions
$\psi(t)$ only on the interval ${\cal I}=(-\frac{\pi}2,\frac{\pi}2)$ without imposing boundary
conditions, the value of $t_0\in \R$ makes a difference. It allows us to pick a segment of length $\pi$
anywhere on the profile of the bounce, not necessarily including its minimum.

\smallskip

\noindent {\bf Explicit form of the Yang--Mills fields.} 
Let us display the explicit non-Abelian Yang--Mills solutions on dS$_4$
corresponding to (\ref{3.15})-(\ref{3.16}) after substituting (\ref{3.5}) and (\ref{3.9}) into (\ref{3.1})
and~(\ref{3.2}). For $\psi=\pm1$ we obtain the trivial solutions $\Fcal\equiv 0$ (vacua). For $\psi=0$ we get the nontrivial
smooth configuration
\begin{subequations}\label{3.19}
\begin{eqnarray}
\Acal \!\!&=&\!\! \sfrac12\,e^a\,I_a \=
\sfrac{1}{2R}\cos t\,\tilde{e}^a\,I_a \= \frac{1}{2R\cosh\tau}\,\tilde{e}^a\,I_a\ ,\\[4pt]
\Fcal \!\!&=&\!\! -\sfrac14\,\ve^a_{bc}\,e^b{\wedge}e^c\,I_a \=
-\sfrac{1}{4R^2}\cos^2\!t\,\ve^a_{bc}\,\tilde{e}^b{\wedge}\tilde{e}^c\,I_a \=
-\frac{1}{4R^2\cosh^2\!\tau}\,\ve^a_{bc}\,\tilde{e}^b{\wedge}\tilde{e}^c\,I_a\ ,
\end{eqnarray}
\end{subequations}
where
\beq\label{3.20}
\tilde{e}^0\ :=\ \frac{R}{\cos t}\,\diff t \= R\,\diff\tau
\qquad\und\qquad
\tilde{e}^a\ :=\ \frac{R}{\cos t}\,e^a \= R\,\cosh \tau\,e^a
\eeq
constitutes an orthonormal basis for the left-invariant one-forms on de Sitter space.
From (\ref{3.19}b) we read off the color-electric and color-magnetic components 
\beq
\Et_a\=\frac{\cos^2\!t}{R^2}\, E_a \und \Bt_a\=\frac{\cos^2\!t}{R^2}\,B_a
\eeq
as
\beq\label{3.21}
\tilde{E}_a\=\tilde{\Fcal}_{0a}\=0 \qquad\und\qquad
\tilde{B}_a\=\sfrac12\ve_{abc}\tilde{\Fcal}_{bc}
\= -\frac{\cos^2\!t}{2R^2}\, I_a=-\frac{1}{2R^2\cosh^2\!\tau}\, I_a\ .
\eeq
with respect to the orthonormal basis (\ref{3.20}), where we used (\ref{2.14}).
In Appendix~D we display the explicit coordinate dependence of these field components 
in different coordinates of~$S^3$.

Inserting the bounce solution (\ref{3.16}), we obtain a family of nonsingular Yang--Mills configurations,
\begin{subequations}\label{3.22}
\begin{eqnarray}
\Acal \!\!&=&\!\! \frac{\cos t}{2R}\,\biggl\{
1\ +\ \frac{\sqrt{2}}{\cosh\bigl(\sqrt{2}(t{-}t_0)\bigr)}\biggr\}\,\tilde{e}^a\,I_a\ ,\\[4pt]
\Fcal \!\!&=&\!\! -\frac{\cos^2\!t}{4R^2}\,\biggl\{
4\frac{\sinh\bigl(\sqrt{2}(t{-}t_0)\bigr)}{\cosh^2\bigl(\sqrt{2}(t{-}t_0)\bigr)}\,
\tilde{e}^0{\wedge}\tilde{e}^a\ +\
\frac{\cosh\bigl(2\sqrt{2}(t{-}t_0)\bigr)-3}{\cosh^2\bigl(\sqrt{2}(t{-}t_0)\bigr)}\,\sfrac12
\ve^a_{bc}\,\tilde{e}^b{\wedge}\tilde{e}^c \biggr\} I_a\ ,
\end{eqnarray}
\end{subequations}
depending on $t_0\in\R$. This family carries electric as well as magnetic fields.
Their relative size as a function of~$\tau$ is displayed in Fig.~2.
\begin{figure}[ht!]
\centering
\begin{minipage}{54mm}
\includegraphics[scale=0.55]{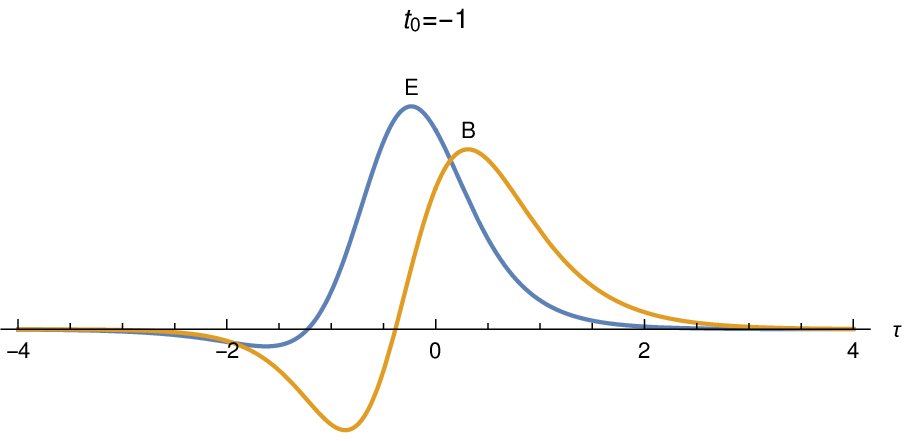}
\\[14pt] $\phantom{.}$
\end{minipage}
\begin{minipage}{54mm}
\includegraphics[scale=0.55]{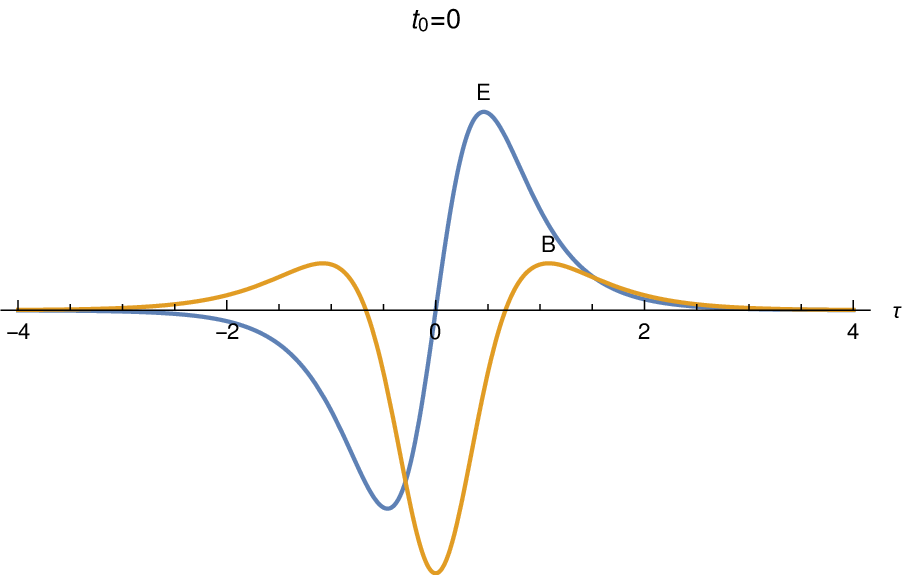}
\end{minipage}
\begin{minipage}{54mm}
\includegraphics[scale=0.55]{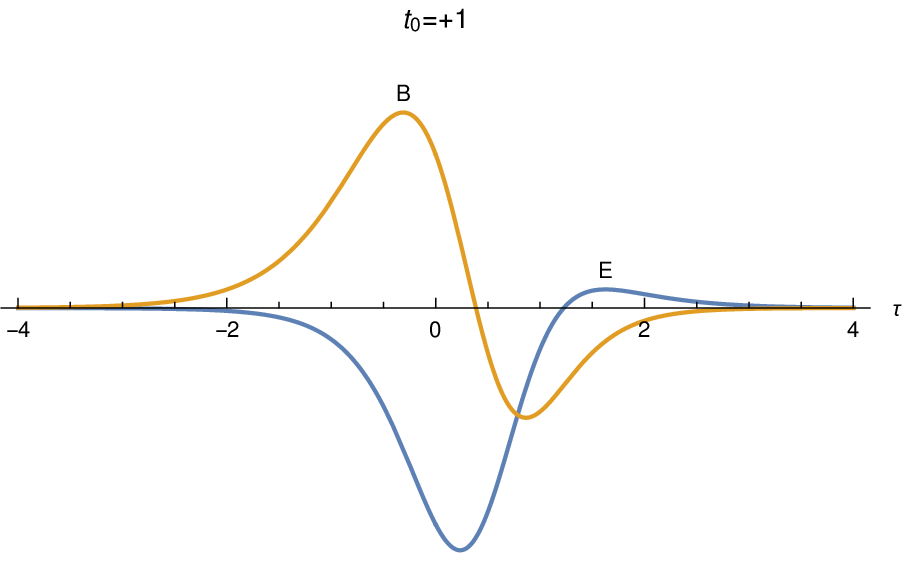} 
\\[-10pt] $\phantom{.}$
\end{minipage}
\caption{Electric and magnetic amplitudes for the bounce configuration (\ref{3.22}b)
with $t_0{=}{-}1,0,{+}1$.}
\label{fig2}
\end{figure}

Finally, for the Abelian solutions, the substitution of (\ref{3.18}) yields
\begin{subequations}\label{3.23}
\begin{eqnarray}
\Acal \!\!&=&\!\! -\sfrac{\ga}{2}\,\cos 2(t{-}t_0)\,{e}^3\,I_3
\=-\sfrac{\ga}{2R}\,\cos t\cos 2(t{-}t_0)\,\tilde{e}^3\,I_3\ ,\\[4pt]
\Fcal \!\!&=&\!\! \diff\Acal 
\= \sfrac{\ga}{R^2}\,\cos^2\!t\,\Bigl\{ \sin{2}(t{-}t_0)\,\tilde{e}^0{\wedge}\tilde{e}^3
\ +\ \cos{2}(t{-}t_0)\,\tilde{e}^1{\wedge}\tilde{e}^2 \Bigr\} I_3\ ,
\end{eqnarray}
\end{subequations}
hence
\begin{subequations}\label{3.24}
\begin{eqnarray}
\Et_3\!\!&=&\!\! \Fct_{03} \= \sfrac{\ga}{R^2}\,\cos^2\!t \,\sin 2(t{-}t_0)\,I_3\ ,\\[4pt]
\Bt_3\!\!&=&\!\! \Fct_{12} \= \sfrac{\ga}{R^2}\,\cos^2\!t \,\cos 2(t{-}t_0)\,I_3\ .
\end{eqnarray}
\end{subequations}
Using (\ref{2.14}), one can rewrite (\ref{3.22})-(\ref{3.24}) in terms of global coordinates $(\tau,\chi,\th,\phi)$ on dS$_4$.

\smallskip

\noindent {\bf Remark.} The Dirac monopole is a connection $a_1$ (see (\ref{2.12})) in the Hopf bundle (\ref{2.11})
over $S^2$, with unit topological charge (1st Chern number) given by (\ref{2.12}). One can embed $S^2$
in $\R^3$ and lift $a_1$ from $S^2$ to $\R^3$. The result is the familiar form of the singular Dirac
monopole solution of the Yang--Mills equations on $\R^3\subset\R^{3,1}$. On the other hand, using the Hopf
fibration (\ref{2.11}) one can pull the monopole connection $a_1$ back to the 3-sphere $S^3$ and obtain
$e^3=\pi^* a_1$. Then, the Abelian gauge connection $\Acal_D=e^3I_3$ on $S^3$ is smooth, but it does not satisfy the
Yang--Mills equations either on $S^3\subset\,$dS$_4$ or on dS$_4$. However, by considering the Abelian
potential $\Acal =\xi(t)\,e^3 I_3$, one obtains the Yang--Mills solution  (\ref{3.18}), (\ref{3.23}) and~(\ref{3.24})
on dS$_4$. It oscillates around the Dirac monopole $\Acal (t{=}t_0)\sim {e}^3\,I_3$ on $S^3$.

\smallskip

\noindent {\bf Energy of the Yang--Mills solutions.}
The energy of Yang--Mills configurations on de Sitter space dS$_4$ computes as
\beq
\begin{aligned}\label{3.25}
\Ecal&\=-\sfrac14\,\int_{S^3(t)}\tilde{e}^1{\wedge}\tilde{e}^2{\wedge}\tilde{e}^3\ \tr(\Et_a\Et_a +\Bt_a\Bt_a) \\
&\=-\frac{\cos t}{4\,R}\int_{S^3}{e}^1{\wedge} {e}^2{\wedge} {e}^3\ \tr(\Fcal_{0a} \Fcal_{0a} + \sfrac12\,
\Fcal_{ab}\Fcal_{ab})\ ,
\end{aligned}
\eeq 
where $S^3(t)$ is the 3-sphere of radius $\frac{R}{\cos t}$. On the configuration (\ref{3.21}), this evaluates to 
\beq\label{3.26} 
\Ecal\=\frac{3\pi^2\,C(j)}{2R\cosh\tau}
\eeq 
after additional use of (\ref{2.14}). 
Similarly, for the configuration (\ref{3.22}) one gets the {\it same\/} result~(\ref{3.26}) 
as for the purely magnetic configuration. The total energy of the Abelian configuration 
(\ref{3.23}) and~(\ref{3.24}) is 
\beq\label{3.27} 
\Ecal_{\textrm{Abelian}}\=\frac{\pi^2\ga^2\,C(j)}{2R\cosh\tau}\ , 
\eeq 
where $\ga^2$ is the moduli parameter.
We see that, for all these gauge-field configurations, the energy decays exponentially for early and late times. 
Its finiteness is quite obvious, since our configurations are non-singular on the finite-volume spatial $S^3$ slices.

\smallskip

\noindent {\bf Action of the Yang--Mills solutions.} 
In a similar fashion one can evaluate the action
functional on the field configurations (\ref{3.19}b), (\ref{3.22}b) and (\ref{3.23}b).
Due to conformal invariance, the action functional can be calculated either in the de Sitter metric (\ref{2.15})
or in the cylinder metric~(\ref{2.16}) on ${\cal I}\times S^3$.
We have
\begin{equation}\label{3.28}
\begin{aligned}
S &\= \sfrac18 \int_{{\rm dS}_4} \tilde{e}^0{\wedge}\tilde{e}^1{\wedge}\tilde{e}^2{\wedge}\tilde{e}^3\
\tr(- 2\Fct_{0a}\Fct_{0a} + \Fct_{ab}\Fct_{ab})\\
&\= \sfrac18 \int_{{\cal I}\times S^3}{e}^0{\wedge}{e}^1{\wedge}{e}^2{\wedge}{e}^3\ \tr(- 2\Fcal_{0a}\Fcal_{0a} +
\Fcal_{ab}\Fcal_{ab}) \ ,
\end{aligned}
\end{equation}
where 
\beq
\Fcal \=\sfrac12\,\Fcal_{\mu\nu}\,e^{\mu}{\wedge}e^{\nu}
\=\sfrac12\,\Fct_{\mu\nu}\,\tilde{e}^{\mu}{\wedge}\tilde{e}^{\nu}
\quad\for \mu,\nu =0,\ldots,3\ ,
\eeq
and the relation between $e^{\mu}$ and $\tilde{e}^{\mu}$ is given in~(\ref{3.20}).

For the purely magnetic configuration (\ref{3.19}b) the action evaluates to
\begin{equation}\label{3.29}
S \= -\sfrac32\pi^3 C(j)\ .
\end{equation}
One may restore the gauge coupling in the denominator.
The action on the `bounce' configuration (\ref{3.22}) comes out as
\begin{equation}\label{3.30}
\frac{S}{C(j)} \= -\sfrac32\pi^3 + 12\,\pi^2\!\!\int_{-\pi/2}^{\pi/2}\!\!\!\!\diff t\,
\frac{\sinh^2\bigl(\sqrt{2}(t{-}t_0)\bigr)}{\cosh^4\bigl(\sqrt{2}(t{-}t_0)\bigr)}
\= -\sfrac32\pi^3 + \sqrt{8}\pi^2
\bigl( \tanh^3(\sfrac{\pi}{\sqrt{2}}{+}\delta) + \tanh^3(\sfrac{\pi}{\sqrt{2}}{-}\delta) \bigr)\ ,
\end{equation}
where $\delta=\sqrt{2}\,t_0\in\R$. Its numerical value varies between 5.52 (for $\delta{=}0$) and -46.51 (for
$\delta\to\pm\infty$). Finally, the action functional on the Abelian solutions (\ref{3.23}) vanish, \beq\label{3.31}
S_{\textrm{Abelian}}\=-\sfrac14\,\int_{{\rm dS}_4} \tilde{e}^0{\wedge}\tilde{e}^1{\wedge}\tilde{e}^2{\wedge}\tilde{e}^3\ \tr(\Et_a\Et_a
-\Bt_a\Bt_a) \=\int_{{\rm dS}_4}\tilde{e}^0{\wedge}\tilde{e}^1{\wedge}\tilde{e}^2{\wedge}\tilde{e}^3\ (\tilde\rho_e -
\tilde\rho_m) \=0\ , \eeq since the integrals of the electric and magnetic energy densities $\tilde\rho_e$ and $\tilde\rho_m$
are finite and equal.

In summary, we have described a class of Abelian and non-Abelian gauge configurations solving
the Yang--Mills equations on de Sitter space dS$_4$. They are spatially homogeneous
and decay for early and late times. Their energies and actions are all finite.

\bigskip

\section{Instantons on de Sitter space dS$_4$}

\noindent
A useful tool to obtain information about the non-perturbative dynamics of gauge theories in flat
space is instanton configurations. It is known that by Euclidean continuation the space
dS$_4$ becomes a 4-sphere $S^4$ of radius $R$ with the metric (\ref{2.21}). Therefore, instantons in
dS$_4$ are the standard $S^4$~instantons. Here we present them in a form adapted
to the coordinates on $S^4$ and the gauge $\Acal_{\vp}=0$.

\smallskip

\noindent {\bf Four-sphere.}
The Euclidean form of the dS$_4$ metric in global coordinates can be obtained by substituting
\begin{equation}
\tau \= \im\,(\vp -\sfrac{\pi}{2}) \qquad\textrm{with}\quad \vp\in [0,\pi ]\ .
\end{equation}
Then the metric (\ref{2.4}) becomes the metric on $S^4$ of radius $R$,
\beq\label{2.21}
\diff s^2 \= R^2(\diff\vp^2 + \sin^2\!\vp\,\diff\Omega^2_3)
\qquad\textrm{with}\quad \diff\Omega^2_3=\delta_{ab}\,e^a e^b\ .
\eeq
This is the standard form in terms of four angles.

By the coordinate transformation
\beq\label{4.4} 
r \= R\,\tan\sfrac{\vp}{2}
\qquad\Longrightarrow\qquad
\sin \vp\=\frac{2Rr}{r^2{+}R^2}\quad,\qquad
\cos\vp\=\frac{R^2{-}r^2}{r^2{+}R^2}
\eeq
it is related to the stereographic coordinates
\beq
x^i \= r\,\omega^i \quad\for i=1,\ldots,4
\quad\with r^2:=\delta_{ij}\,x^i x^j\ ,
\eeq
so that
\beq\label{4.2}
\diff s^2\=\frac{4R^4}{(r^2{+}R^2)^2}\,\de_{ij}\,
\diff x^i\diff x^j \=\frac{4R^4}{(r^2{+}R^2)^2}\,(\diff r^2 + r^2\de_{ab}\,e^a e^b)\ .
\eeq

\smallskip

\noindent {\bf Conformal equivalence of metrics on $S^4$ and $\R{\times}S^3$.} 
The metric (\ref{2.21}) or (\ref{4.2}) is conformally equivalent to the metric on the Euclidean cylinder, 
\beq\label{4.5} 
\diff s^2\= \frac{R^2}{\cosh^2\!T}(\diff T^2+\diff\Omega^2_3) 
\eeq 
via
\beq\label{4.6} 
T \= \frac12\log\frac{1-\cos\vp}{1+\cos\vp} 
\qquad\Longleftrightarrow\qquad 
\ep^T \=\tan\frac{\vp}{2} 
\qquad\Longleftrightarrow\qquad 
\sin\vp\= \frac{1}{\cosh T}
\eeq 
or
\beq\label{4.7} 
T \= \log{\frac{r}{R}}
\qquad\Longleftrightarrow\qquad 
\ep^T\=\frac{r}{R}\ ,
\eeq
respectively.

\smallskip

\noindent {\bf Self-duality.}
The instanton equations on $S^4$,
\beq\label{4.8}
\Fcal_{ij}\=\sfrac12\, \sqrt{\det{g}}\ \ve_{ijkl}\ \Fcal^{kl}\ ,
\eeq
are conformally invariant, and it is more convenient to consider
them on the cylinder $\R\times S^3$ with the metric
\beq\label{4.9}
\diff s^2_{\textrm{cyl}}
\=\diff T^2+\diff\Omega_3^2\=\frac{\cosh^2\!T}{R^2}\,\diff s^2\ .
\eeq
In the basis $(e^i)=(e^a,\diff T)$ the SU(2)-invariant
(spherically symmetric) connection $\Acal$ in the gauge
$\Acal_T=0=\Acal_{\vp}$ and its curvature are given by~\cite{ILPS}
\beq\label{4.10}
\Acal \= X_a \, e^a \quad ,\qquad \
\Fcal_{4 a} \= \frac{\diff X_a}{\diff T}\quad\und\quad
\Fcal_{ab}  \= -2\,\ve_{abc}\,X_c + [X_a , X_b]\ ,
\eeq
and (\ref{4.8}) reduces to a form of the
generalized Nahm equations given by
\beq \label{4.11}
\frac{\diff X_a}{\diff T} \= 2X_a - \sfrac{1}{2}\,\ve_{abc} \, [X_b , X_c]\ .
\eeq
With the same ansatz as previously,
\beq \label{BPS}
X_1=\Psi_1\,I_1\ ,\quad
X_2=\Psi_2\,I_2\ ,\quad
X_3=\Psi_3\,I_3 \quad\with
\Psi_a=\Psi_a(T)\in\R
\eeq
these turn into a coupled set of three ordinary first-order differential equations
(the dot denotes the derivative with respect to~$T$),
\beq \label{flow}
\begin{aligned}
\sfrac12\dot\Psi_1 &\= \Psi_1\ -\ \Psi_2\Psi_3 \= \sfrac12\,\sfrac{\partial U}{\partial\Psi_1}\ ,\\
\sfrac12\dot\Psi_2 &\= \Psi_2\ -\ \Psi_3\Psi_1 \= \sfrac12\,\sfrac{\partial U}{\partial\Psi_2}\ ,\\
\sfrac12\dot\Psi_3 &\= \Psi_3\ -\ \Psi_1\Psi_2 \= \sfrac12\,\sfrac{\partial U}{\partial\Psi_3}
\end{aligned}
\eeq
with the superpotential
\beq\label{superpotential}
U \= \Psi_1^2+\Psi_2^2+\Psi_3^2\ -\ 2\Psi_1\Psi_2\Psi_3\ ,
\eeq
which is depicted in Fig.~3.
\begin{figure}[ht!]
\centering
\includegraphics[scale=1.0]{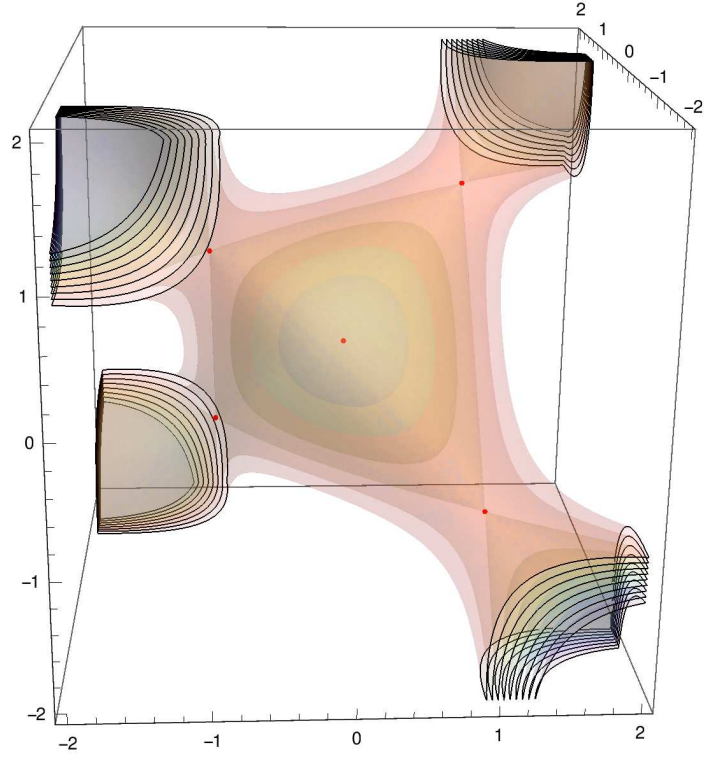}
\caption{Contours of the superpotential potential $U$ in (\ref{superpotential}).}
\label{fig3}
\end{figure}
The corresponding Newtonian dynamics is given by
\beq \label{S4Newton}
\ddot\Psi_a \= +\frac{\partial V}{\partial\Psi_a} 
\quad\with V \= \sfrac12\,\frac{\partial U}{\partial\Psi_a}\,\frac{\partial U}{\partial\Psi_a}
\eeq
yielding the potential given in (\ref{3.7}), but entering with the opposite sign.
Its critical points coincide with the potential minima listed in~(\ref{critpts}),
with values $U(0,0,0)=0$ and $U(1,1,1)=1$.
The flow equations (\ref{flow}) also imply that 
\beq
\sfrac12\dot\Psi_a\dot\Psi_a=V(\Psi) \qquad\und\qquad
\dot U=2\,V\ .
\eeq

\smallskip

\noindent {\bf Instantons.}
The static trajectories $\Psi_a=0$ and $\Psi_a=1$ (and its images under $S_4$) lead to
the trivial vacuum solution $\Fcal=0$. However, there exists an analytic BPS solution
interpolating between the two kinds of critical points of~$U$.
It is captured again by the further simplification
\beq \label{S3ansatz}
\Psi_1=\Psi_2=\Psi_3\=\sfrac12\,(1+\psi) \quad\for \psi=\psi(T)\in\R
\eeq
which leaves us with a single differential equation,
\beq \label{kinkeq}
\dot\psi\=1-\psi^2\=\frac{\partial U_\psi}{\partial\psi}
\quad\with U_\psi \= \psi-\sfrac13\psi^3\ .
\eeq

Its simplest solution is the kink
\beq
\psi(T) \= \tanh(T{-}T_0)
\eeq
with integration constant (or collective coordinate) $T_0$,
which produces
\beq \label{4.12}
X_a\=\bigl[1+\exp(-2(T{-}T_0))\bigr]^{-1}\,I_a
\quad\with [I_a, I_b]=2\,\ve_{ab}^cI_c\ .
\eeq
By using (\ref{4.7}), one can rewrite it as
\beq \label{4.13}
X_a \= \frac{r^2}{r^2{+}\Lambda^2}\,I_a
\quad\for \Lambda^2:=\ep^{2 T_0} R^2\ ,
\eeq
which is exactly the BPST instanton extended from $\R^4$ to
$S^4$~\cite{ILPS}. This is easily seen from
\begin{subequations}\label{4.14}
\begin{eqnarray}
\Acal \!\!&=&\!\! X_ae^a \= -\frac{1}{r^2{+}\Lambda^2}\,\eta^a_{ij}\,I_a\, x^i\diff x^j
\quad\with e^a = -\frac{1}{r^2}\,\eta_{ij}^a\,x^i\,\diff x^j\ ,\\[4pt]
\Fcal \!\!&=&\!\! -\frac{\Lambda^2}{(r^2{+}\Lambda^2)^2}\,\eta^a_{ij} I_a\,\diff x^i\we\diff x^j
\=-\frac{\Lambda^2(r^2{+}R^2)^2}{4R^4(r^2{+}\Lambda^2)^2}\,\eta^a_{ij} I_a\,\tilde{e}^i\we\tilde{e}^j 
\for \tilde{e}^i = \frac{2R^2\,\diff x^i}{r^2{+}R^2}\ ,
\end{eqnarray}
\end{subequations}
where the $\tilde{e}^i$ form an orthonormal basis of one-forms on $S^4$. For $R=\Lambda$ $(T_0=0)$, $\Fcal$ 
has the canonical form of BPST instanton on $S^4$. The radius~$R$ of~$S^4$ sets the scale for~$\Lambda$, but we
may tune $T_0$ in~(\ref{4.13}) such as to remove the dependence of $\Lambda$ on~$R$.
The action of this configuration evaluates to $S=8\pi^2$ independent of~$R$.
The anti-instanton is found by flipping the sign of~$T$.

\smallskip

\noindent {\bf Remark.}
Of course, the Newton equation~(\ref{S4Newton}) has more solutions than the flow equations~(\ref{flow}).
For example, other bounded solutions for $\psi$ oscillate anharmonically between $\psi=-1$ and $\psi=1$. 
When viewed in the full parameter space $\R^3\ni(\Psi_a)$ however, almost all classical trajectories
will run away to infinity, since the inverted potential~$-V$ does not have any local minimum.
This is reflected in the value of the topological charge
\beq
q \= -\frac{1}{64\pi^2 C(j)}\!\int_{\R\times S^3}\!\!\diff T\we e^1\we e^2\we e^3\;
\ve^{ijkl}\,\tr(\Fcal_{ij}\Fcal_{kl})
\= \!\int\!\!\diff T\;\bigl(2(\Psi_1{-}\Psi_2\Psi_3)\dot\Psi_1\,+\,\textrm{cyclic}\bigr)
\= \!\int\!\!\diff T\ \dot U
\eeq
which differs from zero or infinity only if the trajectory $\Psi_a(T)$ connects the two types
of critical points, i.e.~for an instanton or anti-instanton.
Those two saturate the inequality $S\ge4\pi^2 C(j)\,|q|$.

\bigskip

\section{Description of anti-de Sitter space AdS$_4$}

\noindent {\bf AdS$_3$-slicing coordinates.}  
For the remainder of the paper we attempt to repeat the previous analysis
for {\sl anti-\/}de Sitter space AdS$_4$.
In analogy with the dS$_4$ case, where we used the fact that $S^3\cong\,$SU(2) 
is a group manifold, for AdS$_4$ we may employ instead another group manifold,
\begin{equation}\label{5.1}
{\textrm{AdS}}_3\ \cong\ {\textrm{PSL}}(2,\R)\={\textrm{SL}}(2,\R)/\{\pm{\mathbf{1}}_2\}\ ,
\end{equation}
and embed AdS$_4$ into $\R^{3,2}$ in such a way that the metric on AdS$_4$ will be
conformally equivalent to the metric on a cylinder $\R\times\,$PSL$(2,\R)$,
in order to follow our recipe for constructing Yang--Mills solutions.

So, AdS$_4\cong\,$O$(3,2)/$O(3,1) is a hypersurface in $\R^{3,2}$ 
topologically equivalent to $S^1\times\R^3$ and defined by 
\begin{equation}\label{5.2}
(y^1)^2+(y^2)^2-(y^3)^2-(y^4)^2+(y^5)^2 \=-R^2\ .
\end{equation}
One can introduce global coordinates $(z,t,\rho,\phi)$ by setting
\begin{equation}\label{5.3}
y^i = R\,\omega^i\cosh z\ ,\quad y^5 = R\,\sinh z
\quad\with z\in\R \und \eta_{ij}\,\omega^i\omega^j=-1
\end{equation}
for $\omega^i=\omega^i(t,\rho,\phi)$ with $i=1,\ldots,4$ embedding AdS$_3$ into $\R^{2,2}$
with metric $(\eta_{ij})=\textrm{diag}(1,1,{-}1,{-}1)$.
A dimensional coordinate $\tilde z$ can be introduced as $\tilde z=R\,z$.
The flat metric on $\R^{3,2}$ induces a metric on AdS$_4$, 
\beq\label{5.4} 
\diff s^2 \= R^2\,\bigl( \diff z^2 + \cosh^2\!z\,\diff\Omega_{2,1}^2 \bigr)\ , 
\eeq 
where $\diff\Omega_{2,1}^2$ denotes the metric on the unit-radius AdS$_3\cong{\textrm{PSL}}(2,\R)$.

On this space we introduce an orthonormal basis $\{e^\al\}\, ,\ \al=0,1,2,$ of left-invariant one-forms
which satisfy the equations
\begin{equation}\label{5.7}
\diff e^\al + f^\al_{\be\ga}\,e^\be\we e^\ga\=0\ ,
\end{equation}
where
\begin{equation}\label{5.8}
f^\al_{\be\ga}=\eta^{\al\de}\ve_{\de\be\ga} \quad\for
(\eta_{\al\be})=\mbox{diag} ({-}1, {+}1, {+}1)
\end{equation}
are the structure constants of the group SL$(2,\R)$. Concretely, from
(\ref{5.8}) we have
\begin{equation}\label{5.9}
f^1_{20}=f^2_{01}=1 \und f^0_{12}=-1 \quad\for \ve^{}_{012}=1\ .
\end{equation}
In terms of $e^\al$ the AdS$_3$ metric has the form
\begin{equation}\label{5.10}
\diff\Omega_{2,1}^2 \= \eta_{\al\be}\,e^\al e^\be
\= -(e^0)^2+(e^1)^2+(e^2)^2\ .
\end{equation}
Explicit formul\ae\ for coordinates and one-forms on unit AdS$_3$
can be found in Appendix~C.

\smallskip

\noindent 
{\bf Conformal coordinates I.} 
Instead of the coordinate $z$ one can introduce the coordinate
\begin{equation}\label{5.15}
\chi = \arctan (\sinh z)\quad\Longleftrightarrow\quad \sinh
z=\tan\chi\ ,\quad \cosh z=\frac{1}{\cos\chi}
\end{equation}
in which $z\in (-\infty, \infty)$ corresponds to $\chi\in (-\frac{\pi}{2},\frac{\pi}{2})$. 
The metric (\ref{5.4}) in the coordinates $(\chi,t,\rho,\phi)$ reads
\begin{equation}\label{5.16}
\diff s^2 \= \frac{R^2}{\cos^2\!\chi}\,\bigl( \diff \chi^2 + \diff\Omega_{2,1}^2 \bigr) \=
\frac{R^2}{\cos^2\!\chi}\,\diff s^2_{\textrm{cyl}}  \ ,
\end{equation}
where
\begin{equation}\label{5.17}
\diff s^2_{\textrm{cyl}} \= \diff \chi^2 + \eta_{\al\be}\,e^\al e^\be 
\= \eta_{\m\n}\,e^\m e^\n \quad\for \m,\n , =0,\ldots,3 \und e^3:=\diff\chi
\end{equation}
is the metric on the cylinder $\R\,{\times}\,$AdS$_3$ with the Minkowski metric
\begin{equation}\label{5.18}
(\eta_{\m\n}) \= \textrm{diag}(\eta_{\al\be},{+}1) \= \textrm{diag}({-}1,{+}1,{+}1,{+}1)
\end{equation}
in the orthonormal basis $(e^\m)$. Hence, we see that anti-de Sitter space is conformally equivalent to the finite
cylinder ${\cal I}\times\,$PSL$(2,\R)$ with the interval ${\cal I}=(-\frac{\pi}{2},\frac{\pi}{2})$, 
fully parallel to de Sitter space after substituting PSL$(2,\R)\cong\,$AdS$_3$ for of SU(2)$\,\cong S^3$ and
switching the signature of the cylinder coordinate.

\smallskip

\noindent 
{\bf $H^3$-slicing coordinates.} 
Anti-de Sitter space is the one-sheeted hyperboloid embedded in flat $\R^{3,2}$ by the relation~(\ref{5.2}). 
Another natural slicing is provided by the global coordinates $(t,\rho,\th,\phi)$ 
\begin{equation}\label{5.19}
y^1\!=R\sinh\!\rho\,\la^2\,,\
y^2\!=R\sinh\!\rho\,\la^1\,,\
y^3\!=R\cosh\!\rho\,\cos t\,,\
y^4\!=R\cosh\!\rho\,\sin t\,,\
y^5\!=R\sinh\!\rho\,\la^3\,,
\end{equation}
where $t\in [-\pi,\pi)$ parametrizes a circle, $\rho\ge 0$, 
and $\la^a=\la^a(\th,\phi)$ from (\ref{2.18}) embed $S^2$ into $\R^3$ in the standard manner.
The flat metric on $\R^{3,2}$ induces on AdS$_4$ the metric
\begin{equation}\label{5.20}
\diff s^2 = R^2 \bigl(-\cosh^2\!\rho\,\diff t^2 + \diff\rho^2 +\sinh^2\!\rho\,\diff\Omega_2^2\bigr)
\quad\with \diff\Omega_2^2\=\diff\th^2 +\sin^2\!\th\diff\phi^2\ ,
\end{equation}
showing that equal-time slices are hyperbolic 3-spaces~$H^3$.
A dimensionful time coordinate $\tilde t$ can be introduced as $\tilde t=R\,t$ with $\tilde t\in [-\pi R, \pi R)$. 
The conformal boundary $\rho\to \infty$ for this metric has topology $S^1\times S^2$ with coordinates $(t,\th,\phi)$. 
One can unwrap the circle $S^1=\R/\Z$ and extend the time coordinate~$t$ to all of~$\R$, which means considering
the universal covering space $\widetilde{\mbox {AdS}}_4$ of AdS$_4$ having topology $\R^4$ instead of $S^1\times\R^3$.

\smallskip

\noindent 
{\bf Conformal coordinates II.} 
Instead of the coordinate $\rho$ in (\ref{5.19}) and (\ref{5.20}) one can introduce the coordinate~\cite{PH}\footnote{
This coordinate $\chi$ differs from $\chi$ in (\ref{5.15}) but agrees with (half of) $\chi$ parametrizing $S^3$ in (\ref{A.2}).}
\begin{equation}\label{5.21}
\chi = \arctan (\sinh\rho )\quad\Longleftrightarrow\quad 
\sinh\rho = \tan\chi\ ,\quad \cosh\rho=\frac{1}{\cos\chi}\ ,
\end{equation}
in which $\rho\in [0,\infty)$ corresponds to $\chi\in [0,\sfrac{\pi}{2})$. 
The metric (\ref{5.20}) in the coordinates $(t,\chi,\th,\phi)$ reads
\begin{equation}\label{5.22}
\diff s^2 \= \frac{R^2}{\cos^2\!\chi}\bigl(-\diff t^2 +\diff\chi^2+\sin^2\!\chi\,\diff\Omega_2^2\bigr)
\= \frac{R^2}{\cos^2\!\chi} \bigl(-\diff t^2 + \diff\Omega_{3+}^2\bigr) 
\end{equation}
where $\diff\Omega_{3+}^2$ is the metric on the upper hemisphere $S^3_+$ of the 3-sphere $S^3=S^3_+\cup S^2\cup S^3_-$, 
since the conformal boundary $\rho\to\infty$ has been retracted to the finite boundary at $\chi{=}\sfrac{\pi}{2}$
corresponding to the equator of~$S^3$ for any value of~$t$. 
The metric $\diff\Omega_{3+}^2$ differs from the $S^3$ metric in~(\ref{A.2}) only by the range of~$\chi$
($\sfrac{\pi}{2}$ rather than $\pi$).
Both patches $S^3_\pm$, introduced in (\ref{2.10}), have the topology of~$\R^3$. 
However, the metric on $S^3_+$ is the standard metric on the 3-sphere and can be written as
\begin{equation}\label{5.23}
\diff \Omega_{3+}^2\=\de_{ab}\,e^ae^b\ ,
\end{equation}
where the $e^a$ are defined in (\ref{2.5})-(\ref{2.7}) but are considered only on the upper hemisphere. 

{}From (\ref{5.22}) we see that the metric on anti-de Sitter space is also conformally equivalent to
\begin{equation}\label{5.24}
\diff s^2_{\textrm{cyl}}\=-\diff t^2 + \de_{ab}\,e^a e^b \=\eta_{\mu\nu}\,e^\mu e^\nu
\quad\for \mu,\nu=0,\ldots,3 \und e^0:=\diff t\ ,
\end{equation}
where
\beq \label{tranges}
t\in[-\pi,\pi) \for \textrm{AdS}_4 \qquad\und\qquad t\in\R \for \widetilde{\rm AdS}_4\ ,
\eeq
so we are dealing with $S^1{\times}S^3_+\cong S^1{\times}\R^3$ or with a Lorentzian cylinder $\R\times S^3_+\cong\R^4$, respectively.
One should be clear about which space, AdS$_4$ or ${\widetilde{\rm AdS}}_4$, is considered.

\bigskip

\section{Yang--Mills configurations on AdS$_4$}

\noindent
Here we describe some Yang--Mills configurations on AdS$_4$ with the metric (\ref{5.16}) conformally equivalent to the
cylinder metric (\ref{5.17}) on ${\cal I}\times\textrm{AdS}_3$ and group structure on AdS$_3\cong\textrm{PSL}(2,\R)$ discussed in
Section 5. Of course, the list of solutions we construct is not exhaustive. First we will consider solutions which are
naturally described in the metrics (\ref{5.16}) and (\ref{5.17}). Their energy and action are infinite due to infinite volume
of the space AdS$_3\cong\textrm{PSL}(2,\R)$. Then we will find solutions naturally described in the metric
(\ref{5.22})-(\ref{5.24}) and will show that both their energy and action are {\sl finite\/} on AdS$_4$.

Similar to Section~3, solutions on AdS$_4$ can be obtained by solving the Yang--Mills equations on ${\cal I}\times\textrm{PSL}(2,\R)$
with the metric (\ref{5.17}) or on $S^1{\times}S^3_+$ with the metric (\ref{5.24}). And again, on SU($N$)-valued gauge-group
elements $g(\cdot)$, acting on $su(N)$-valued gauge fields $\Acal$ and $\Fcal$, we can impose the boundary condition 
$g(\pa({\cal I}\times\textrm{AdS}_3))=\Id$ on the boundary 
$\pa({\cal I}\times\textrm{AdS}_3)=\textrm{AdS}_3|_{\chi=\pm\pi/2}=\textrm{AdS}_3|_{z=\pm\infty}$ and similarly $g=\Id$ 
on the boundary $\pa({\cal T}\times S^3_+)={\cal T}\times S^2$ for the metric (\ref{5.24}), 
where ${\cal T}=S^1$ for AdS$_4$ and  ${\cal T}=\R$ for ${\widetilde{\rm AdS}}_4$.

\smallskip

\noindent {\bf Matrix equations.}  
We first employ the metrics (\ref{5.16}) and (\ref{5.17}).  We will be concise in our
discussion since all our steps will repeat those from Section 3. For $su(N)$-valued gauge potentials $\Acal$ in the gauge
$\Acal_3=\Acal_\chi =0=\Acal_z$ we employ the ansatz
\begin{equation}\label{6.1}
\Acal \= X_\al (\chi ) \, e^\al\ .
\end{equation}
Here, $X_\al (\chi )$ are three $su(N)$-valued functions depending
only on $\chi\in\cal I$, and $e^\al$ are one-forms on PSL$(2,\R)$
given in (\ref{5.6}) and satisfying (\ref{5.7}). The
field strength for this ansatz reads
\begin{equation}\label{6.2}
{\cal F} \= \Fcal_{3\al}\, e^3\we e^\al + \sfrac12\Fcal_{\al\be}\, e^\al\we e^\be 
\= X'_\al\, e^3\we e^\al + \sfrac12(-2f^\al_{\be\ga} X_{\al} + [X_\be , X_\ga ])\, e^\be\we e^\ga \ ,
\end{equation}
where $X'_\al :=\diff X_\al /\diff\chi$ and 
$e^\mu{=}(e^\al,e^3){=}(e^\al,\diff\chi )$ for $\al =0,1,2$. 
After substitution of (\ref{6.1}) and (\ref{6.2}), the Yang--Mills equations
on ${\cal I}\times\,$PSL$(2,\R)$ reduce to the matrix differential equations
\begin{equation}\label{6.3}
\begin{aligned}
X''_0 &\= -4X_0\ -\ 6[X_1,X_2]\ +\ \bigl[X_1,[X_0,X_1]\bigr] + \bigl[X_2,[X_0,X_2]\bigr]\  ,\\
X''_1 &\= -4X_1\ +\ 6[X_2,X_0]\ +\ \bigl[X_2,[X_1,X_2]\bigr] - \bigl[X_0,[X_1,X_0]\bigr]\  ,\\
X''_2 &\= -4X_2\ +\ 6[X_0,X_1]\ -\ \bigl[X_0,[X_2,X_0]\bigr] + \bigl[X_1,[X_2,X_1]\bigr]\  ,
\end{aligned}
\end{equation}
where $X''_\al :=\diff^2X_\al/\diff\chi^2$. Comparison to (\ref{3.3}) shows that the two sets
of equations are related by\footnote{
The sign choice for $X_1$ and for $X_2$ must be the same.}
\beq \label{su2sl2}
X_3\ \mapsto\ X_0\ ,\qquad X_1\ \mapsto\ \pm\im\,X_1\ ,\qquad X_2\ \mapsto\ \pm\im\,X_2\ ,
\eeq
reflecting the relation between the SU(2) and SL(2,$\R$) generators.

\smallskip

\noindent
{\bf Reduction to particle mechanics.}  
As in Section~3, we take the matrices $X_\al$ from a spin-$j$ representation of SU(2)
with generators $(I_1,I_2,I_3)$ inside $su(N)$ with $N=2j{+}1$ and put
\begin{equation}\label{6.4}
X_0=\Psi_0\,I_3\ ,\quad X_1=\Psi_1\,I_1 \und X_2=\Psi_2\,I_2\ ,
\end{equation}
where $\Psi_\al$ are real functions of $\chi$. Substituting
(\ref{6.4}) into (\ref{6.3}), we obtain
\begin{equation}\label{6.5}
\begin{aligned}
\sfrac14 \Psi_0'' &\= -\Psi_0 - 3\,\Psi_1\Psi_2 + \Psi_0(\Psi_1^2+\Psi_2^2)
\= -\sfrac{\pa V}{\pa\Psi_0}\ ,\\
\sfrac14 \Psi_1'' &\= -\Psi_1 + 3\,\Psi_2\Psi_0 - \Psi_1(\Psi_0^2-\Psi_2^2)
\= +\sfrac{\pa V}{\pa\Psi_1}\ ,\\
\sfrac14 \Psi_2'' &\= -\Psi_2 + 3\,\Psi_0\Psi_1 - \Psi_2(\Psi_0^2-\Psi_1^2)
\= +\sfrac{\pa V}{\pa\Psi_2}
\end{aligned}
\end{equation}
for a quasi-potential function
\beq \label{6.8}
V \= 2\,\bigl\{ (\Psi_0+\Psi_1\Psi_2)^2-(\Psi_1-\Psi_2\Psi_0)^2-(\Psi_2-\Psi_0\Psi_1)^2\bigr\}\ ,
\eeq
consistent with~(\ref{su2sl2}).
However, the interpretation of a Newtonian dynamics is disturbed by the fact that the
quasi-kinetic energy
\beq
T \= \sfrac12\,(\dot\Psi_0^2-\dot\Psi_1^2-\dot\Psi_2^2) \= -\sfrac12\,\eta^{\al\be}\,\dot\Psi_\al\dot\Psi_\be
\eeq
inherits the indefiniteness of the AdS$_3$ metric, giving a negative `mass' to $\Psi_1$ and~$\Psi_2$.

The Yang--Mills Lagrangian on the cylinder ${\cal I}\times\textrm{PSL}(2,\R)$ becomes
\begin{equation}\label{6.7}
{\cal L} \= \sfrac18 \, \tr\,\Fcal_{\m\n}\Fcal^{\m\n} 
\= 2\,C(j)\,(T - V)\ ,
\end{equation}
and the Euler--Lagrange equations derived from (\ref{6.7}) coincide with (\ref{6.5}). 

\smallskip

\noindent {\bf Solutions.} 
The system (\ref{6.5}) is invariant only under a $D_8$ subgroup of the tetrahedral symmetry of~(\ref{3.8}).
Therefore, the Abelian dS$_4$ solution (\ref{3.10}) also applies here,
\beq \label{6.10}
\Psi_1=\Psi_2=0 \und \Psi_0=\xi \qquad\Rightarrow\qquad
\xi'' \= -4\,\xi \qquad\Rightarrow\qquad
\xi(\chi) \= \sfrac{\ga}{2}\,\cos 2(\chi{-}\chi_0)\ ,
\eeq
where $\ga$ and $\chi_0$ are arbitrary real parameters.
Analogous solutions exist exciting only $\Psi_1$ or $\Psi_2$.

We cannot write down nontrivial analytic solutions of the system (\ref{6.5}).
In particular, there is no analog of the bounce solution (\ref{3.16}) to (\ref{3.9}) and~(\ref{3.12}).
However, static solutions exist, since the quasi-potential (\ref{6.8}) has the 5 critical points
$(\hat\Psi_0,\hat\Psi_1,\hat\Psi_2)$ with values
\beq\label{6.13}
(0,0,0) \ \Rightarrow\ V=0 \quad,\qquad
(\pm3,\pm1,\pm1) \ \Rightarrow\ V=16\quad,
\eeq
where the number of minus signs in each triple must be even.
The $\Psi_\al{=}0$ configuration corresponds to the vacuum solution~$\Fcal{=}0$,
while the other critical points yield genuine non-Abelian Yang--Mills solutions.

\smallskip

\noindent {\bf Yang--Mills solutions with infinite action.}
Let us display the explicit form of Yang--Mills configurations on AdS$_4$
corresponding to the solutions (\ref{6.10}) and (\ref{6.13}) of the reduced
Yang--Mills equations (\ref{6.5}).
Computing their energy and action entails integrating over the spatial part of the
AdS$_3$ slice, which is the hyperbolic space~$H^2$. Since the latter has infinite volume,
as can be seen from the metric~(\ref{5.5}), the energy and the action of these solutions
are infinite.

Substituting $(\Psi_0,\Psi_1,\Psi_2)=(3,1,1)$ into (\ref{6.4}), (\ref{6.1}) and (\ref{6.2}), 
we obtain the solution
\begin{equation}\label{6.18}
\Acal \= \sfrac{1}{R} \cos\chi\,\bigl(3\,I_3\,\tilde{e}^0+I_1\,\tilde{e}^1+I_2\,\tilde{e}^2\bigr) \und
\Fcal \= \sfrac{4}{R^2} \cos^2\!\chi\,
\bigl( 2\,I_3\,\tilde{e}^1\we\tilde{e}^2+I_2\,\tilde{e}^0\we\tilde{e}^1+I_1\,\tilde{e}^2\we\tilde{e}^0 \bigr)\ ,
\end{equation}
where  $\tilde{e}^\m = R\,e^\m{/}\cos\chi$ for $(\mu)=(\al,3)$ 
is the orthonormal basis on AdS$_4$ for the metric~(\ref{5.16}).
We read off the color-electric and color-magnetic components
\begin{equation}\label{6.19}
\begin{aligned}
\Et_1 &\= \Fct_{01}\=\sfrac{4}{R^2\cosh^2\!z}\,I_2\ , \\
\Et_2 &\= \Fct_{02}\=\sfrac{-4}{R^2\cosh^2\!z}\,I_1\ , \\
\Bt_3 &\= \Fct_{12}\=\sfrac{8}{R^2\cosh^2\!z}\,I_3\ ,
\end{aligned}
\end{equation}
where we used the relations (\ref{5.15}). All other components vanish since $\Fct_{\al 3}=0$. 
Flipping two of the signs in the solution $(\Psi_0,\Psi_1,\Psi_2)$ produces analogous configurations,
which differ from (\ref{6.18}) and~(\ref{6.19}) only by switching the signs of two in three terms
correspondingly.

Substituting (\ref{6.10}) into (\ref{6.4}), (\ref{6.1}) and (\ref{6.2}),
we get the Abelian solution
\begin{subequations}\label{6.15}
\begin{eqnarray}
\Acal \!&=&\! \sfrac{\ga}{2}\,\cos 2(\chi{-}\chi_0)\,{e}^0\,I_3 \=
\sfrac{\ga}{2R}\,\cos\chi\,\cos 2(\chi{-}\chi_0)\,\tilde{e}^0\,I_3\ ,\\[4pt]
\Fcal \!&=&\! \diff\Acal \= \sfrac{\ga}{R^2}\,\cos^2\!\chi\,\bigl\{
\sin{2}(\chi{-}\chi_0)\,\tilde{e}^0\we\tilde{e}^3 +
\cos{2}(\chi{-}\chi_0)\,\tilde{e}^1\we\tilde{e}^2 \bigr\}\,I_3
\end{eqnarray}
\end{subequations}
and therefore
\begin{subequations}\label{6.16}
\begin{eqnarray}
\Et_3\!&=&\! \Fct_{03} \= \sfrac{\ga}{R^2}\,\cos^2\!\chi \,\sin 2(\chi{-}\chi_0)\,I_3\ ,\\[4pt]
\Bt_3\!&=&\! \Fct_{12} \= \sfrac{\ga}{R^2}\,\cos^2\!\chi \,\cos 2(\chi{-}\chi_0)\,I_3\ .
\end{eqnarray}
\end{subequations}
Using the correspondence (\ref{5.15}) one can rewrite (\ref{6.15}) and
(\ref{6.16}) in terms of the $z$~coordinate used in~(\ref{5.4}). 
For the Abelian solution, the action is proportional to
\begin{equation}\label{6.17}
\mbox{vol}\,(\mbox{PSL} (2,\R))\times \int^{\pi/2}_{-\pi/2}\!\!\diff\chi\ 
\bigl(\sin^2\!2(\chi{-}\chi_0) - \cos^2\!2(\chi{-}\chi_0)\bigr)\ .
\end{equation}
The above integral vanishes but it multiplies the infinite group volume.\footnote{
We may regularize the volume of AdS$_3$ before integrating over $\chi$ and thus obtain a vanishing action in this case.}

\smallskip

\noindent {\bf Yang--Mills solutions with finite action.} 
Now we consider the metric (\ref{5.22}) on AdS$_4$.
Thanks to the conformal invariance of the Yang--Mills equations in four dimensions, 
it suffices to study Yang--Mills theory on $S^1{\times}S^3_+$ with the metric (\ref{5.24})
including the standard metric (\ref{A.2}) restricted to the upper hemisphere~$S^3_+$.
Since the one-forms $e^a$ in (\ref{5.23}) obey (\ref{2.5}), we can literally copy the ansatz (\ref{3.1}) 
for ${\cal I}\times S^3$ to our space $S^1\times S^3_+$,
\beq\label{3.1'}
\Acal = X_a(t)\,e^a\ .
\eeq
Then all formul\ae\ (\ref{3.3})-(\ref{3.15}) are valid in this case as well, yielding the same
matrix equations, three-dimensional Newtonian dynamics and its solutions as in the de Sitter case.
However, the periodicity in~$t$ in addition requires
\beq
\Psi_a(t{+}2\pi) = \Psi_a(t)\ .
\eeq
Consequently, the constant and periodic Yang--Mills solutions (\ref{3.19}) and (\ref{3.23}) on dS$_4$
are also valid on AdS$_4$, after changing their conformal factor,
\beq \label{6.20}
\tilde{e}^\mu = \frac{R}{\cos t}\,e^\mu \qquad\Rightarrow\qquad
\tilde{e}^\mu = \frac{R}{\cos\chi}\,e^\mu \qquad\with e^0=\diff t\ ,
\eeq
and restricting $e^a$ to the upper hemisphere~$S^3_+$.
The bounce solution (\ref{3.16}) does not qualify. However, it is the limiting case 
of a continuum of periodic solutions given by a Jacobi elliptic function,
\begin{equation}\label{6.20a}
\psi(t;k)\=\alpha(k)\,{\rm dn}[\alpha(k)(t{-}t_0);k] \quad\with 
\alpha(k)\=\sqrt{2/(2{-}k^2)} \und 0\le k< 1\ .
\end{equation}
This family interpolates between the bounce (\ref{3.16}) for $k\to1$ (period $\to\infty$) and
the constant vacuum solution $\psi=1$ for $k=0$. For infinitesimally small values of~$k$ we
have harmonic oscillations with a period of~$\pi$, as can be gleaned from a harmonic approximation to ~(\ref{3.12}).
By continuity, shorter periods cannot be attained, but there should exist a $2\pi$-periodic solution for a special value of~$k$.
Since dn$[u;k]$ has a period of $2{\cal K}(k)$, where ${\cal K}(k)$ is the complete elliptic integral of the first kind 
(see e.g.\ the Appendix of \cite{BILL} for a brief discussion of Jacobi functions), 
the periodicity condition is satisfied if \cite{MaS, LMKT}
\begin{equation}\label{6.21a}
\frac{{\cal K}(k)}{\al(k)} \= \pi \qquad\Rightarrow\qquad k=\bar{k}\approx 0.9977\ ,
\end{equation}
which is very near to the bounce, as can be seen in Fig.~4.
\begin{figure}[ht!]
\centering
\includegraphics[scale=1.0]{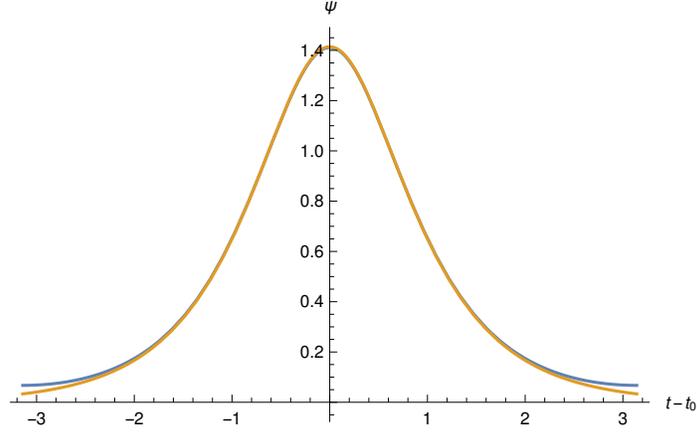}
\caption{Bounce (yellow) and $2\pi$-periodic (blue) solution in the double-well potential (\ref{3.12}).}
\label{fig4}
\end{figure}

We lift all these solutions from the cylinder (\ref{5.24}) to AdS$_4$ with metric (\ref{5.22}) written as
\begin{equation}\label{6.21}
\diff s^2 \=\frac{R^2}{\cos^2\!\chi}\,(-\diff t +\diff\Omega_{3+}^2 ) \= \eta_{\mu\nu}\,\te^\mu\te^\nu
\end{equation}
and obtain
\begin{subequations}\label{6.22}
\begin{eqnarray}
\Acal \!\!&=&\!\! \sfrac12\,e^a\,I_a \=
\sfrac{1}{2R}\cos \chi\,\tilde{e}^a\,I_a\ ,\\[4pt]
\Fcal \!\!&=&\!\! -\sfrac14\,\ve^a_{bc}\,e^b{\wedge}e^c\,I_a \=
-\sfrac{1}{4R^2}\cos^2\!\chi\,\ve^a_{bc}\,\tilde{e}^b{\wedge}\tilde{e}^c\,I_a 
\end{eqnarray}
\end{subequations}
for the colour-magnetic solution,
\begin{subequations}\label{6.23}
\begin{eqnarray}
\Acal \!\!&=&\!\!\sfrac{1}{2\,R}\,\cos\chi\,\bigl(1+\al(\bar k)\,{\rm dn}[\ldots]\bigr)\,\tilde{e}^a\,I_a\ ,\\[4pt]
\Fcal \!\!&=&\!\! \sfrac{1}{2\,R^2}\,\cos^2\!\chi\,\Bigl\{
\bar k\,\al(\bar k)^2\,{\rm cn}[\ldots]\,{\rm sn}[\ldots]\,\tilde{e}^0{\wedge}\tilde{e}^a -
\sfrac12\,\bigl(1{-}\al(\bar k)^2\,{\rm dn}^2[\ldots]\bigr)\,\ve^a_{bc}\,\tilde{e}^b{\wedge}\tilde{e}^c \Bigr\}\,I_a
\end{eqnarray}
\end{subequations}
for the near-bounce solution, with arguments $[\ldots]=[\al(\bar k)(t{-}t_0)]$ 
for Jacobi elliptic functions cn, sn and dn, and
\begin{subequations}\label{6.24}
\begin{eqnarray}
\Acal \!\!&=&\!\! -\sfrac{\ga}{2}\,\cos 2(t{-}t_0)\,{e}^3\,I_3
\=-\sfrac{\ga}{2\,R}\,\cos\chi\,\cos 2(t{-}t_0)\,\tilde{e}^3\,I_3\ ,\\[4pt]
\Fcal \!\!&=&\!\! \diff\Acal \= \frac{\ga}{R^2}\,\cos^2\!\chi\,
\bigl\{ \sin{2}(t{-}t_0)\,\tilde{e}^0{\wedge}\tilde{e}^3\ +\ \cos{2}(t{-}t_0)\,\tilde{e}^1{\wedge}\tilde{e}^2 \bigr\}\,I_3
\end{eqnarray}
\end{subequations}
for the Abelian solution which stems from (\ref{3.18}). Using (\ref{5.21}),
one can rewrite (\ref{6.22})-(\ref{6.24}) in terms of coordinates $(t,\rho ,\th ,\phi )$ on AdS$_4$.

\smallskip

\noindent{\bf Energy of the Yang--Mills solutions.} 
The energy of Yang--Mills configurations on anti-de Sitter space AdS$_4$ computes as 
\beq
\begin{aligned}\label{6.25}
\Ecal&\=-\sfrac14\,\int_{\tilde{S}^3_+}\tilde{e}^1{\wedge}\tilde{e}^2{\wedge}\tilde{e}^3\ 
\tr(\Fct_{0a}\Fct_{0a} + \sfrac12\,\Fct_{ab}\Fct_{ab}) \\
&\=-\frac{1}{4\,R}\int_{S^3_+}{e}^1{\wedge} {e}^2{\wedge} {e}^3\ 
\cos\chi\ \tr(\Fcal_{0a} \Fcal_{0a} + \sfrac12\, \Fcal_{ab}\Fcal_{ab})\ ,
\end{aligned}
\eeq 
where $\tilde{S}^3_+$ is the hemisphere with its metric conformally rescaled by $\frac{R^2}{\cos^2\!\chi}$.

On the configuration (\ref{6.22}), the energy evaluates to
\begin{equation}\label{6.27}
\Ecal \= \frac{\pi\,C(j)}{R}\ .
\end{equation}
We see that it is not only finite but, contrary to the de Sitter case, it does not depend on time. Hence, we get a static
magnetic configuration. For the configuration (\ref{6.23}) we use results about sphalerons on a circle \cite{MaS, LMKT} to
calculate the energy and find
\begin{equation}\label{6.27a}
\Ecal(k) \= \frac{\pi\,C(j)}{R}\,\frac{4\,k^2}{(1+k^2)^2}\ ,
\end{equation}
which correctly interpolates between the values for the vacuum ($k=0$) and the static magnetic solution ($k\to1$).
For the admissible value of~$k$, its value $\Ecal(\bar k)$ differs from (\ref{6.27}) by a factor of about $1-5{\times}10^{-6}$.
Finally, the energy of the Abelian configuration (\ref{6.24}) is
\begin{equation}\label{6.28}
\Ecal_{\textrm{Abelian}} \= \frac{\pi\,\ga^2\,C(j)}{3\,R}\ ,
\end{equation}
where $\ga^2$ is the moduli parameter. So, for all three Yang--Mills solutions the energy is finite and constant.

\smallskip

\noindent{\bf Action of the Yang--Mills solutions.} 
As previously, the action functional on the field configurations (\ref{6.22})-(\ref{6.24})
can be calculated either in the anti-de Sitter metric (\ref{5.22}) or in the cylinder metric (\ref{5.24}) on
$S^1{\times}S^3_+$, 
\begin{equation}\label{6.29}
\begin{aligned}
S &\= \sfrac18 \int_{{\rm AdS}_4} \tilde{e}^0{\wedge}\tilde{e}^1{\wedge}\tilde{e}^2{\wedge}\tilde{e}^3\
\tr(- 2\Fct_{0a}\Fct_{0a} + \Fct_{ab}\Fct_{ab})\\
&\= \sfrac18 \int_{S^1{\times}S^3_+}{e}^0{\wedge}{e}^1{\wedge}{e}^2{\wedge}{e}^3\ \tr(- 2\Fcal_{0a}\Fcal_{0a} +
\Fcal_{ab}\Fcal_{ab}) \ .
\end{aligned}
\end{equation}

For the purely magnetic configuration (\ref{6.22}) the action evaluates to
\begin{equation}\label{6.30}
S \= -\sfrac32\,\pi^3 C(j)\ .
\end{equation}
Interestingly, it coincides with the value (\ref{3.29}) for the analogous but non-static configuration on~dS$_4$. 
This is because they are identical but are lifted from different spaces which happen to have the same volume,
\beq
\textrm{vol}({\cal I}{\times}S^3) \= \pi\times 2\pi^2 \= 2\pi\times\pi^2 \= \textrm{vol}(S^1{\times}S^3_+) \ .
\eeq
The action of the configuration (\ref{6.23}) is reduced to 
\begin{equation}\label{6.31}
S(k) \= \sfrac34\,\pi^2 C(j) \int_{S^1}\!\diff t\ \Bigl\{
\dot\psi(t;k)^2\ -\ \bigl(1-\psi(t;k)^2\bigr)^2 \Bigr\} \quad\for k=\bar k\ ,
\end{equation}
where $\psi(t;k)$ is periodic and given in (\ref{6.20a}).  
This integral is finite and independent of~$t_0$ but cannot be written down analytically.
Its numerical value is about 41\%\ of (\ref{6.30}).
Finally, the action functional on the Abelian solution~(\ref{6.24}) vanishes,
\begin{equation}\label{6.32}
S_{\textrm{Abelian}}\=0\ ,
\end{equation}
because the integral of the electric and magnetic energy densities are finite and equal.

\smallskip

\noindent{\bf Boundary values of the Yang--Mills solutions.}
Since anti-de Sitter space has a boundary, it is of interest to note the value our solutions take there.
The infinite-action field components (\ref{6.19}) and (\ref{6.16}) as well as the finite-action fields
(\ref{6.22}b), (\ref{6.23}b) and (\ref{6.24}b) all carry the conformal factor $\cosh^{-2}\!z=\cos^2\!\chi$,
which vanishes at the boundary $z\to\pm\infty$ or $\chi=\pm\sfrac{\pi}{2}$. Therefore, our solutions
live in the subspace of gauge fields decaying to zero at the AdS$_4$ boundary.

\bigskip

\section{Self-dual Yang--Mills fields on anti-de Sitter space AdS$_4$}

\noindent {\bf Euclidean AdS$_4$.} 
Finally we discuss instantons in anti-de Sitter space AdS$_4$. The Euclidean continuation of
the AdS$_3$ metric (\ref{5.5}) is obtained by substituting $t=\im\tau$,
which turns the $z{=}\textrm{const}$ slices to 3-dimensional hyperbolic spaces~$H^3$.
The metric on AdS$_4$ transforms to a $\cosh$-cone metric on the hyperbolic space $H^4$. This form
of metric on $H^4$ is not convenient for our study of instantons since the natural boundary of $H^4$ 
is the 3-sphere $S^3=\pa H^4$. 
However, there exist various other choices of coordinates and metrics on AdS$_4$ (see e.g.~\cite{ale}),
such as
\begin{subequations}\label{7.1}
\begin{eqnarray}
\diff s^2 \!\!&=&\!\! R^2 \bigl(-\cosh^2\!\rho\,\diff t^2 +\diff\rho^2 + \sinh^2\!\rho\,\diff\Omega_2^2\bigr) \ ,\\[4pt]
\diff s^2 \!\!&=&\!\! R^2 \bigl( - \diff t^2 + \sin^2\!t\,(\diff\rho^2 + \sinh^2\!\rho\,\diff\Omega_2^2)\bigr)\ ,\\[4pt]
\diff s^2 \!\!&=&\!\! R^2 \bigl(\diff\rho^2 + \sinh^2\!\rho\,(-\diff t^2 + \cosh^2\!t\,\diff\Omega_2^2)\bigl)\ .
\end{eqnarray}
\end{subequations}
Choosing  $t=\im (\chi{-}\sfrac{\pi}{2})$ in (\ref{7.1}c) one obtains for $H^4$ a $\sinh$-cone metric
over~$S^3$,
\begin{equation}\label{7.6}
\diff s^2 \= R^2 \bigl(\diff\rho^2 + \sinh^2\!\rho\,(\diff\chi^2+\sin^2\!\chi\,\diff\Omega_2^2)\bigr)
\= R^2(\diff\rho^2 + \sinh^2\!\rho\,\de_{ab}\,e^ae^b)\ ,
\end{equation}
which is convenient for analyzing gauge instantons on AdS$_4$.\footnote{
Also convenient~\cite{ABT} is the Euclidean continuation of (\ref{7.1}a). Instantons in AdS$_4$ and
${\widetilde{\rm AdS}}_4$ with this metric in the form (\ref{5.22}) will be considered at the end of this section.} 
Here, $e^a$ are the left-invariant one-forms on $S^3$ satisfying (\ref{2.5}) and discussed in detail in Appendix~B.
We remark that, due to the range $\rho>0$ the metric~(\ref{7.6}) describes only one sheet of the two-sheeted hyperboloid
in~$\R^{4,1}$ as a complete model of Euclideanized AdS$_4$.
Furthermore, we must eventually fix some boundary conditions for the gauge fields, 
in order to investigate stability, for instance. 
Here, we take the attitude to postpone this discussion and first learn about classical solutions
for any kind of boundary condition.

\smallskip

\noindent{\bf Cylinder metric.} 
In stereographic coordinates $x^i$, $i=1,\ldots,4$, the $H^4$ metric reads
\begin{equation}\label{7.7}
\diff s^2 \= \frac{4\,R^4}{(r^2{-}R^2)^2}\,\de_{ij}\,\diff x^i\diff x^j
\for r^2 = \de_{ij}\,x^i x^j\ <\ R^2\ ,
\end{equation}
which resembles the metric~(\ref{4.2}) on $S^4$.
The forms (\ref{7.7}) and~(\ref{7.6}) are related by the coordinate transformation
\begin{equation}\label{7.8}
r\=R\,\tanh\frac{\rho}{2} \quad\with r\in [0, R) \quad\Leftrightarrow\quad\rho\in[0,\infty )\ .
\end{equation}
Further, the metric (\ref{7.6}) is conformally equivalent to the metric (\ref{4.9}) on the Euclidean cylinder,
\begin{equation}\label{7.9}
\diff s^2 \= R^2(\diff\rho^2+\sinh^2\!\rho\,\diff\Omega_3^2)
\=\frac{R^2}{\sinh^2\! T}(\diff T^2+\diff\Omega^2_3)
\=\frac{R^2}{\sinh^2\! T}\,\diff s^2_{\textrm{cyl}}\ ,
\end{equation}
where
\begin{equation}\label{7.10}
T \=\log\tanh\sfrac{\rho}{2}\qquad\Longleftrightarrow\qquad
\tanh\sfrac{\rho}{2}\=\ep^T\qquad\Longleftrightarrow\qquad
\sinh\rho \=\frac{1}{\sinh T}\ .
\end{equation}

\smallskip

\noindent{\bf BPST-type quasi-instanton.} 
The Yang--Mills self-duality equations (\ref{4.8}) are valid on any four-manifold. For
the metric (\ref{7.9}) on $H^4$ they are reduced to the equations on the cylinder $\R\times S^3$ with the metric (\ref{4.9})
and become the generalized Nahm equations~(\ref{4.11}) for three matrices $X_a$. 
Therefore, we can copy the kink solution presented in Section~4,
\beq \label{7.11}
\Acal \= X_a\,e^a \quad,\qquad
X_a \= \sfrac12(1+\psi)\,I_a \quad,\qquad
\psi(T) \= \tanh(T{-}T_0)\ ,
\eeq
where the $I_a$ are defined by (\ref{3.4}) and $T_0$ is a real parameter.
Thus we see that up to this moment the analysis of the self-dual Yang--Mills equations 
is the same on $\R^4$ (as a metric-cone over $S^3$), 
on $S^4$ (as a sine-cone over $S^3$), or on $H^4$ (as a sinh-cone over $S^3$).
The differences appear only in the range of $T$ and in the role of moduli parameter~$T_0$.

First, for the cylinder $\R{\times}S^3$ we have $T\in (-\infty ,\infty)$, yielding 
\beq\label{7.14} 
\Acal (T {=}{-}\infty)\=0 \quad\und\quad \Acal (T{=}{+}\infty)\=e^a I_a\=g^{-1}\diff\,g\ , 
\eeq 
where $g(\chi,\th ,\phi): S^3\to\textrm{SU}(2)$ is a smooth map of degree (winding number) one. 
Thus, $\Acal(T)$ describes a transition from the trivial vacuum (sector of topological charge $q{=}0$) to a
nontrivial vacuum (sector $q{=}1$). 
Second, for the sphere~$S^4$ one takes $T\in[-\infty,\infty]$, corresponding to $\vp\in [0,\pi]$. 
Hence, the self-dual solution again has topological charge $q{=}1$ and extends the one from $\R^4$ to~$S^4$. 
In more detail, for the gauge field from (\ref{7.11}) we get 
\beq\label{7.16} 
\Fcal \= -\dot{X}_a\, e^a\we e^4+\sfrac12\,\bigl(-2\ve^a_{bc} X_a +[X_b,X_c]\bigr)\,e^b\we e^c 
\= \frac{1}{4\cosh^2(T{-}T_0)}\,\eta^a_{ij}\,e^i\we e^j I_a\ , 
\eeq 
where $e^4:=\diff T$.  It follows that 
\beq\label{7.17} 
q\ :=\ -\frac{1}{16\pi^2 C(j)}\int_{\R\times S^3} \tr (\Fcal\we\Fcal ) \= 1\ .
\eeq 
This integral depends neither on the metric nor on~$T_0$ and is the same 
for $\R^4$, $\R{\times}S^3$ and $S^4$.

Third, turning to hyperbolic space $H^4$, we see that 
\beq\label{7.18} 
r\in [0,R) \qquad\Longleftrightarrow\qquad 
\rho\in [0,\infty) \qquad\Longleftrightarrow\qquad
T\in [-\infty,0) \ . 
\eeq 
This means that our solution (\ref{7.11}) and~(\ref{7.16}) is defined only on the {\sl half\/} line~${\R_-}$ 
and describes a transition 
\beq\label{7.19} 
\begin{aligned}
\textrm{from} \qquad
&\Acal(T{=}{-}\infty)\=0 \qquad\textrm{to}\qquad
\Acal(T{=}0)\=\sfrac12(1{-}\tanh T_0)\,e^aI_a \ ,\\[4pt]
\textrm{i.e.} \qquad
&\Fcal(T{=}{-}\infty)\=0 \qquad\textrm{to}\qquad
\Fcal(T{=}0)\=\frac{1}{4\cosh^2\!T_0}\,\eta^a_{ij}\,e^i\we e^j I_a\ ,
\end{aligned}
\eeq 
connecting the trivial vacuum with an instanton section of size $\Lambda=\ep^{T_0}R$, as discussed in~\cite{CW}.
Its quasi-topological charge depends on the moduli parameter~$T_0$,
\beq\label{7.22}
q(T_0)\=-\frac{1}{16\pi^2C(j)} \int_{\R_-\times S^3} \tr (\Fcal\we\Fcal ) 
\= \frac{3\,\ep^{-T_0}+\ep^{-3T_0}}{8\,\cosh^3\!T_0}
\eeq 
ranging from $q{=}0$ for $T_0{\to}\infty$ to $q{=}1$ for $T_0{\to}{-}\infty$.
For $T_0{=}0$ the boundary configuration sits in the middle of the kink, and so the solution (\ref{7.11}) and~(\ref{7.16})
on~$\R_-$ corresponds to a meron~\cite{AFF} (a singular non-self-dual Yang--Mills solution) 
which has topological charge $q=\sfrac12$ in agreement with (\ref{7.22}). 
We also see that for self-dual configurations on $\R_-{\times}S^3$ (and hence on $H^4$) the action functional 
$S=4\pi^2C(j)\,|q|$ decreases monotonically with $T_0$. 

\smallskip

\noindent{\bf Geometric quasi-instanton.} 
When studing instantons on AdS$_4$, one more possibility opens up. The simple flow equation~(\ref{kinkeq}) 
has, besides the kink in~(\ref{7.11}), also the singular solution 
\beq\label{7.23} 
\psi(T) \= \coth (T{-}T_0)\ . 
\eeq 
It must be discarded on $\R^4$ and on $S^4$ due to the pole at $T=T_0$. However, for $T_0>0$ there is no
singularity on the domain $[-\infty,0)\ni T$ relevant for~$H^4$.
Substituting (\ref{7.23}) into (\ref{7.11}), we obtain the self-dual solution
\beq\label{7.24} 
\Acal\=\sfrac12\,\bigl(1+\coth(T{-}T_0)\bigr)\,e^aI_a \quad\und\quad 
\Fcal\=\frac{1}{4\sinh^2(T{-}T_0)}\, \eta^a_{ij}\,e^i\we e^j\, I_a \ .
\eeq 
With (\ref{4.7}) and (\ref{4.13}) we get 
\beq\label{7.25} 
\frac{1}{4\sinh^2(T{-}T_0)}\=\frac{\Lambda^2\,r^2}{(r^2{-}\Lambda^2)^2}\ , 
\eeq
and (\ref{7.24}) coincides with the self-dual Yang--Mills configuration on $H^4$ naturally appearing in the geometric
construction of~\cite{Popov}, for example.\footnote{
On $S^4{=}\,$Sp(2)$/$Sp(1)$\times$Sp(1) the instanton is naturally described as the self-dual part of the Levi-Civita 
connection in the fibration Sp(2)$/$Sp(1)$\ \to S^4$. Analogously, (\ref{7.24}) is the self-dual part of the Levi-Civita 
connection on $H^4{=}\,$Sp(1,1)$/$Sp(1)$\times$Sp(1), which is a connection in the fibration Sp(1,1)$/$Sp(1)$\ \to H^4$. Here, 
Sp(1,1) is the non-compact subgroup of Sp(2)$\otimes\C$ preserving the indefinite metric diag(1,-1) on the quaternionic space $\Hb^2$.}
The topological charge of (\ref{7.24}) comes out as
\beq\label{7.26} 
q(T_0)\=-\frac{1}{16\pi^2C(j)} \int_{\R_-\times S^3} \tr (\Fcal\we\Fcal ) 
\= \frac{3\,\ep^{-T_0}-\ep^{-3T_0}}{8\,\sinh^3\!T_0}\ ,
\eeq 
where now $T_0> 0$.
Thus, just like (\ref{7.16}), the solution (\ref{7.24}) has finite action.

\smallskip

\noindent{\bf Instantons in ${\widetilde{\rm AdS}}_4$.} 
For a fuller picture we take a look at self-dual solutions on the universal cover~${\widetilde{\rm AdS}}_4$.
To this end we perform a Euclidean continuation of the metric~(\ref{5.22}) as proposed in~ \cite{ABT},
\begin{equation}\label{7.27}
\diff s^2 \= \frac{R^2}{\cos^2\!\chi}\bigl(\diff T^2 +\diff\chi^2+\sin^2\!\chi\,\diff\Omega_2^2\bigr)
\= \frac{R^2}{\cos^2\!\chi} \bigl(\diff T^2 + \diff\Omega_{3+}^2\bigr)
\quad\with \chi\in [0,\sfrac{\pi}{2})\ ,
\end{equation}
where $T\in S^1$ for AdS$_4$ but $T\in\R$ for ${\widetilde{\rm AdS}}_4$. 
As before, $S^3_+$ denotes the upper hemisphere, with a volume of~$\pi^2$.
The conformal boundary of this metric corresponds to $\chi=\sfrac{\pi}{2}$ and has the topology 
of a Euclidean space $S^1\times S^2$ or $\R\times S^2$, respectively.

As before, the self-duality equations reduce to equations on the cylinder, but over $S^3_+$ instead of $S^3$.
After taking the canonical ansatz (\ref{4.10}) and specializing to (\ref{BPS}) and~(\ref{S3ansatz}) 
we again obtain the flow equation~(\ref{kinkeq}).
However, this equation admits no periodic solution, so we do not find BPS configurations on Euclideanized AdS$_4$
in this way. 
On the other hand, the Euclildean version of the universal cover ${\widetilde{\rm AdS}}_4$ relaxes the periodicity
requirement. Therefore, on this space we can take the canonical kink solution~(\ref{7.11}) defined for $T\in\R$.
Then the gauge field~(\ref{7.16}) has the unit topological charge~(\ref{7.17}). 
Thus, standard instantons are well defined on the Euclidean version of the universal covering ${\widetilde{\rm AdS}}_4$  
of anti-de Sitter space.

Non-self-dual Yang--Mills solutions on Euclideanized AdS$_4$ can nevertheless be found.
The trusted ansatz (\ref{4.10}), (\ref{BPS} and~(\ref{S3ansatz}) reduces the full Yang--Mills equations to
\begin{equation}\label{7.29}
\frac{\diff^2\psi}{\diff T^2} \= -2\,\psi\,(1-\psi^2)\ ,
\end{equation}
whose periodic solutions in terms of Jacobi elliptic functions are described e.g.~in~\cite{ILPR}. 
Substituting back into the ansatz yields non-self-dual finite-action Yang--Mills configurations,
which describe a sequence of instanton-anti-instanton pairs.

\bigskip

\section{Conclusions}

\noindent
We have established the existence of solitonic classical pure Yang--Mills configurations 
with finite energy and action in four-dimensional de Sitter and anti-de Sitter spaces. 
No Higgs fields are required. 
On de Sitter space dS$_4$ described as spatial $S^3$ slices over real time,
our Yang--Mills solutions are spatially homogeneous and decay exponentially for early and late times. 
Replacing $S^3$ with AdS$_3$ yields infinite-action configurations on AdS$_4$.
However, on anti-de Sitter space AdS$_4$ parametrized as spatial $H^3$ slices over a temporal circle,
we again constructed solutions having finite energy and action,  
which decay exponentially in the radial direction of the hyperbolic slices.\footnote{
In the construction we strongly employed the conformal equivalence of $H^3$ to a 3-hemisphere $S^3_+$,
see (\ref{5.20})-(\ref{tranges}).}
For the Euclideanized version of $\R^{3,1}$, dS$_4$ and AdS$_4$, our method reproduces the known 
BPST instantons and lifts them from $\R^4$ to $S^4$ and $H^4$, respectively, where in the latter case
we get only `half' the instanton.

Due to their finite action, the described gauge configurations should be relevant 
in a semiclassical analysis of the path integral for quantum Yang--Mills theory on dS$_4$ or AdS$_4$.
Their existence indicates that the Yang--Mills vacuum structure may depend on the cosmological constant,
and the question of their stability calls for a computation of the (one-loop) effective action around 
these field configurations. One might hope to employ the (anti-)de Sitter radius~$R$ as a regulator
towards quantum Yang--Mills theory on Minkowski space.

The most symmetric solution has an elementary geometric dependence on de Sitter time
or on anti-de Sitter radial distance, and its action in both cases takes the minimal value of $-3\pi^3$ 
(for the SU(2) adjoint representation with normalization (\ref{3.4})), independent the (anti-)de Sitter 
radius. We conjecture this to be a lower bound for Yang--Mills solutions on these backgrounds. 
It will be important to investigate the stability of our configurations for certain boundary conditions.

Our solutions derive from three simplifying ans\"atze. First, we restricted the gauge potential to 
an $su(2)$ subalgebra and made an SU(2)-equivariant ansatz, which turns the Yang--Mills equations into
ordinary coupled differential equations for three matrices. Second, we took these matrices to be
proportional to the SU(2) generators, which produces a 3-dimensional Newtonian dynamical system with
tetrahedral ($S_4$) symmetry. Third, we focus on stable submanifolds in the parameter space, which 
enables us to find analytic solutions.

At each step, generalizations are possible. First, one may admit a larger gauge group and
a more general ansatz for the matrices, which will lead to quiver gauge theories. 
Second, it is tempting to analyze the matrix dynamics directly, for the potential and superpotential
\bea
&&V \= -\tr\bigl\{ 2\,X_a X_a\ -\ \ve_{abc}\,X_a\,[X_b,X_c]\ +\ \sfrac12\,[X_a,X_b]\,[X_a,X_b] \bigr\}\ ,\\[4pt]
&&U \= -\tr\bigl\{ X_a X_a\ -\ \sfrac16\,\ve_{abc}\,X_a\,[X_b,X_c] \bigr\}\ ,
\eea
respectively.
And third, for a good understanding of the analog Newtonian system one should investigate also
numerical solutions for its full 3-dimensional dynamics. We hope to address these issues in the near future.

\bigskip
\noindent {\bf Acknowledgements}

\noindent
This work was partially supported by the Deutsche Forschungsgemeinschaft grant LE~838/13
and by the Heisenberg-Landau program.
It is based upon work from COST Action MP1405 QSPACE, 
supported by COST (European Cooperation in Science and Technology).

\newpage

\appendix

\section{Four-dimensional metrics used in this paper}

\noindent
{\bf Metrics on dS$_4$.} 

\begin{tabular}{|lcl|}
\hline\vphantom{\Big|}
$\diff s^2/R^2$ & \qquad coordinates \qquad\qquad & range \\[4pt]
\hline\vphantom{\Big|}
$-\diff\tau^2+\cosh^2\!\tau\,\diff\Omega_3^2$ & 
$(\tau,\chi,\th,\ph)$ & $\tau\in\R$ \\[6pt]
$\cos^{-2}\!t\,\bigl(-\diff t^2+\diff\Omega_3^2\bigr)$ & 
$(t,\chi,\th,\ph)$ & $t\in(-\sfrac{\pi}{2},\sfrac{\pi}{2})$ \\[6pt]
$-(1{-}\rho^2)\,\diff\sigma^2+\sfrac{\diff\rho^2}{1{-}\rho^2}+\rho^2\diff\Omega_2^2$ & 
$(\sigma,\rho,\th,\ph)$ & $\sigma\in\R$, $\rho\in[0,1)$ \\[4pt]
\hline
\end{tabular}

\bigskip

\noindent
{\bf Metrics on $S^4$.}

\begin{tabular}{|lcl|}
\hline\vphantom{\Big|}
$\diff s^2/R^2$ & \qquad coordinates \qquad\qquad & range \\[4pt]
\hline\vphantom{\Big|}
$\diff\vp^2+\sin^2\!\vp\,\diff\Omega_3^2$ & 
$(\vp,\chi,\th,\ph)$ & $\vp\in[0,\pi]$ \\[6pt]
$\sfrac{4\,R^2}{(r^2{+}R^2)^2}\,\bigl(\diff r^2+r^2\,\diff\Omega_3^2\bigr)$ & 
$(r,\chi,\th,\ph)$ & $r\in\R_+$ \\[6pt]
$\cosh^{-2}\!T\,\bigl(\diff T^2+\diff\Omega_3^2\bigr)$ & 
$(T,\chi,\th,\ph)$ & $T\in\R$ \\[4pt]
\hline
\end{tabular}

\bigskip

\noindent
{\bf Metrics on AdS$_4$.}

\begin{tabular}{|lcl|}
\hline\vphantom{\Big|}
$\diff s^2/R^2$ & \qquad coordinates \qquad\qquad & range \\[4pt]
\hline\vphantom{\Big|}
$\diff z^2+\cosh^2\!z\,\diff\Omega_{2,1}^2$ &
$(z,t,\rho,\ph)$ & $z\in\R$ \\[6pt]
$\cos^{-2}\!\chi\,\bigl(\diff\chi^2+\diff\Omega_{2,1}^2\bigr)$ &
$(\chi,t,\rho,\ph)$ & $\chi\in(-\sfrac{\pi}{2},\sfrac{\pi}{2})$ \\[6pt]
$-\cosh^2\!\rho\,\diff t^2+\diff\rho^2+\sinh^2\!\rho\,\diff\Omega_2^2$ &
$(t,\rho,\th,\ph)$ & $t\in[-\pi,\pi]$, $\rho\in\R_+$ \\[6pt]
$\cos^{-2}\!\chi\,\bigl(-\diff t^2+\diff\Omega_{3+}^2\bigr)$ &
$(t,\chi,\th,\ph)$ & $t\in[-\pi,\pi]$ \\[6pt]
$-\diff t^2+\sin^2\!t\,\bigl(\diff\rho^2+\sinh^2\!\rho\,\diff\Omega_2^2\bigr)$ &
$(t,\rho,\th,\ph)$ & $t\in[-\pi,\pi]$ , $\rho\in\R_+$ \\[6pt]
$\diff\rho^2+\sinh^2\!\rho\,\bigl(-\diff t^2+\cosh^2\!t\,\diff\Omega_2^2\bigr)$ &
$(\rho,t,\th,\ph)$ & $t\in\R$, $\rho\in\R_+$ \\[4pt]
\hline
\end{tabular}

\bigskip

\noindent
{\bf Metrics on $H^4$.}

\begin{tabular}{|lcl|}
\hline\vphantom{\Big|}
$\diff s^2/R^2$ & \qquad coordinates \qquad\qquad & range \\[4pt]
\hline\vphantom{\Big|}
$\diff\rho^2+\sinh^2\!\rho\,\diff\Omega_3^2$ &
$(\rho,\chi,\th,\ph)$ & $\rho\in\R_+$ \\[6pt]
$\sfrac{4\,R^2}{(r^2{-}R^2)^2}\,\bigl(\diff r^2+r^2\,\diff\Omega_3^2\bigr)$ &
$(r,\chi,\th,\ph)$ & $r\in[0,R)$ \\[6pt]
$\sinh^{-2}\!T\,\bigl(\diff T^2+\diff\Omega_3^2\bigr)$ &
$(T,\chi,\th,\ph)$ & $T\in\R_-$ \\[6pt]
$\cos^{-2}\!\chi\,\bigl(\diff T^2+\diff\Omega_3^2\bigr)$ &
$(T,\chi,\th,\ph)$ & $\chi\in[0,\sfrac{\pi}{2})$ \\[4pt]
\hline
\end{tabular}

\newpage

\section{Metrics on $S^3$}

\noindent 
A standard embedding of $S^3$ into $\R^4$ is given by 
\beq \label{2.3} 
\om^1=\sin\ch\,\sin\th\,\sin\ph\ ,\quad
\om^2=\sin\ch\,\sin\th\,\cos\ph\ ,\quad 
\om^3=\sin\ch\,\cos\th\ ,\quad 
\om^4=\cos\ch\ , 
\eeq 
where $0\le\ch,\th\le\pi$ and $0\le\ph<2\pi$. 
It induces on $S^3$ the metric 
\beq\label{A.2} 
\diff\Omega_3^2 \= \diff\ch^2+\sin^2\!\ch\,(\diff\th^2+\sin^2\!\th\,\diff\ph^2)\ . 
\eeq
For $0\le\chi <\sfrac{\pi}{2}$ this is the metric on the 3-ball $S^3_+$, 
for $\sfrac{\pi}{2}<\chi\le \pi$ it is the metric on the 3-ball $S^3_-$, 
and for $\chi=\sfrac{\pi}{2}$ we have the equatorial~$S^2$, 
in the decomposition $S^3=S^3_+\cup S^2\cup S^3_-$.
Employing (\ref{2.6})
the corresponding one-forms $\{e^a\}$ read 
{\small 
\beq
\begin{aligned}
e^1 \! &\= \sin\th\sin\ph\,\diff\ch
+ \sin\ch\cos\ch\,(\tan\ch\cos\ph{+}\cos\th\sin\ph)\,\diff\th
+ \sin^2\!\ch\sin\th\,(\cot\ch\cos\ph{-}\cos\th\sin\ph)\,\diff\ph \,,\\
e^2 \! &\= \sin\th\cos\ph\,\diff\ch
- \sin\ch\cos\ch\,(\tan\ch\sin\ph{-}\cos\th\cos\ph)\,\diff\th
- \sin^2\!\ch\sin\th\,(\cot\ch\sin\ph{+}\cos\th\cos\ph)\,\diff\ph \,,\\
e^3 \! &\= \cos\th\,\diff\ch
- \sin\ch\cos\ch\sin\th\,\diff\th
+ \sin^2\!\ch\sin^2\!\th\,\diff\ph\,,
\end{aligned}
\eeq
}
in terms of which the metric reads
\beq
\diff\Omega_3^2 \= \bigl(e^1\bigr)^2+\bigl(e^2\bigr)^2+\bigl(e^3\bigr)^2\ .
\eeq

A simpler expression for $\{e^a\}$ arises from the different embedding choice
\begin{equation}\label{2.8}
\om^1=\cos\chi\cos\sfrac{\th}{2}\ ,\quad
\om^2=-\sin\chi\cos\sfrac{\th}{2}\  ,\quad
\om^3=\cos(\phi{-}\chi)\sin\sfrac{\th}{2}\ ,\quad
\om^4=\sin(\phi{-}\chi)\sin\sfrac{\th}{2}\ ,
\end{equation}
where the angles $0\le\th \le\pi$ and $0\le\phi, \chi\le 2\pi$ differ from those used
above (but are denoted the same).  For (\ref{2.8}) substituted in (\ref{2.6}) one obtains
\beq
\begin{aligned}\label{2.9}
e^1 &\= \sfrac12\bigl(\sin(2\chi{-}\phi)\,\diff\th-\cos(2\chi{-}\phi)\sin\th\,\diff\ph\bigr)\ ,\\
e^2 &\= \sfrac12\bigl(\cos(2\chi{-}\phi)\,\diff\th+\sin(2\chi{-}\phi)\sin\th\,\diff\ph\bigr)\ ,\\
e^3 &\=  \sfrac12\bigl(\diff(2\chi{-}\phi)+\cos\th\,\diff\ph\bigr) 
\= \diff\chi-\sfrac12(1{-}\cos\th)\,\diff\phi\ .
\end{aligned}
\eeq Correspondingly, the induced metric on the unit 3-sphere reads
\beq\label{A.5}
\diff\Omega_3^2 \= \de_{ab}\,e^ae^b
\= \bigl(\diff\chi -\sfrac12(1{-}\cos\th)\,\diff\phi\bigr)^2 +
\sfrac14\bigl(\diff\th^2+\sin^2\!\th\,\diff\phi^2\bigr)\ . 
\eeq
This metric is adapted to the Hopf fibration 
\beq\label{2.11} 
\pi\, : \quad S^3 \ \stackrel{U(1)}{\longrightarrow} \ S^2 
\eeq 
with the one-monopole connection\footnote{ 
This is the form on the patch of $S^2$ around $\th{=}0$.
Around $\th{=}\pi$ one should take $a_1=\sfrac{\im}{2}\,(1+\cos\th\,\diff\phi )$.} 
\beq\label{2.12} 
a_1=-\sfrac{\im}{2}\,(1-\cos\th\,\diff\phi)\qquad\Rightarrow\qquad 
f_1=\diff a_1=-\sfrac{\im}{2}\,\sin\th\,\diff\th\we\diff\phi 
\quad\und\quad \sfrac{\im}{2\pi}\smallint_{S^2} f_1=1
\eeq 
entering the metric (\ref{A.5}).

\bigskip

\section{Metrics on AdS$_3$}

\noindent
A standard embedding of AdS$_3$ into $\R^{2,2}$ is given by
\begin{equation}
\om^1=\sinh\rho\,\cos\ph\ ,\quad
\om^2=\sinh\rho\,\sin\ph\ ,\quad
\om^3=\cosh\rho\,\cos t\ ,\quad 
\om^4=\cosh\rho\,\sin t\ ,
\end{equation}
where $-\pi\le t < \pi$, $\rho\ge 0$ and $0\le \phi < 2\pi$. 
It induces on AdS$_3$ the metric
\begin{equation}\label{5.5}
\diff\Omega_{2,1}^2 \= -\cosh^2\!\rho\,\diff t^2 + \diff\rho^2 + \sinh^2\!\rho\,\diff\phi^2\ .
\end{equation}

One can introduce an orthonormal basis $\{e^\al\}$ of left-invariant one-forms 
\beq
\begin{aligned}\label{5.6}
e^0 &\= -\cosh^2\!\rho\,\diff t-\sinh^2\!\rho\,\diff\phi\ ,\\
e^1 &\= \sin(t{-}\phi)\,\diff\rho - \sinh\rho\,\cosh\rho\,\cos(t{-}\phi)\,\diff(t{+}\phi)\ ,\\
e^2 &\= -\cos(t{-}\phi)\,\diff\rho - \sinh\rho\,\cosh\rho\,\cos(t{-}\phi)\,\diff(t{+}\phi)\ ,
\end{aligned}
\eeq 
in terms of which the metric reads
\beq
\diff\Omega_{2,1}^2 \= -\bigl(e^1\bigr)^2+\bigl(e^2\bigr)^2+\bigl(e^3\bigr)^2\ .
\eeq

\bigskip

\section{Yang--Mills solutions on dS$_4$ in various coordinates}

\noindent
We have constructed pure SU(2) Yang--Mills solutions in dS${}_4$ parametrized
by classical double-well trajectories $\psi(t)$. 
It is remarkable that their action is finite. Their scale is set by the inverse de Sitter radius $R^{-1}$. 
Here we display these solutions in different coordinates on de Sitter space.

\subsection{Yang--Mills configuration in closed slicing}

The solutions found in~\cite{IvLePo} for the closed slicing and described in more detail in Section 3
depend on a suitable function $\psi(\tau)$ and have the form
\beq
\Acal \= \sfrac12\, (1{+}\psi)\,e^a I_a  \quad\und\quad
\Fcal \= \bigl(\sfrac12\,\sfrac{\diff\psi}{\diff\tau}\,\diff\tau\we e^a
-\sfrac14\,(1{-}\psi^2)\,\ve^a_{bc}\,e^b\we e^c\bigr)I_a\ ,
\eeq
with three SU(2) generators $\{I_a\}$ and left-invariant one-forms $\{e^a\}$ obeying 
\beq
[I_b, I_c] \= 2\,\ve^a_{bc}\,I_a \ ,\qquad
\tr(I_a I_b) \= -4\,C(j)\,\de_{ab}\ ,\qquad
\diff e^a + \ve^a_{bc}\,e^b\wedge e^c\=0\ ,
\eeq
where $C(j)=\sfrac13\,j(j{+}1)(2j{+}1)$ is the second-order Dynkin index of the spin-$j$ representation.

For extracting the components of $\Acal$ and $\Fcal$, it is convenient to define three matrices $I_*$ via
\beq \label{Imatrices}
e^a I_a \ =:\ \diff\ch\,I_\ch+\diff\th\,I_\th+\diff\ph\,I_\ph \ ,
\eeq
from which it follows that
\beq
\sfrac12\ve^a_{bc}\,e^b{\wedge}e^c\,I_a \=
-\sfrac1{\sin\th}\,\diff\ch{\wedge}\diff\th\,I_\ph
+\sin\th\,\diff\ch{\wedge}\diff\ph\,I_\th
-\sin^2\!\ch\sin\th\,\diff\th{\wedge}\diff\ph\,I_\ch\ .
\eeq
In the fundamental representation of SU(2), $I_a=-\im\,\sigma_a$ and $C(\sfrac12)=\sfrac12$, and so from
\beq
e^a I_a \= -\im \biggl( \begin{matrix}
e^3 & e^1{-}\im e^2 \\[4pt]
e^1{+}\im e^2 & -e^3
\end{matrix} \biggr) \quad\und\quad
\sfrac12\ve^a_{bc}\,e^b{\wedge}e^c\,I_a \= -\im \biggl( \begin{matrix}
e^{12} & e^{23}{-}\im e^{31} \\[4pt]
e^{23}{+}\im e^{31} & -e^{12}
\end{matrix} \biggr)
\eeq
we compute
\beq
\begin{aligned}
I_\ch &\= -\im\, \biggl( \begin{matrix}
\cos\th & {-}\im\sin\th\,\ep^{\im\ph} \\[4pt]
\im\sin\th\,\ep^{-\im\ph} & -\cos\th
\end{matrix} \biggr) \ , \\[6pt]
I_\th &\= -\im\,\sin\ch\cos\ch\, \biggl( \begin{matrix}
-\sin\th & (\tan\ch{-}\im\cos\th)\ep^{\im\ph} \\[4pt]
(\tan\ch{+}\im\cos\th)\ep^{-\im\ph} & \sin\th
\end{matrix} \biggr) \ , \\[6pt]
I_\ph &\= -\im\,\sin^2\!\ch\sin\th\, \biggl( \begin{matrix}
\sin\th & (\cot\ch{+}\im\cos\th)\ep^{\im\ph} \\[4pt]
(\cot\ch{-}\im\cos\th)\ep^{-\im\ph} & -\sin\th
\end{matrix} \biggr) \ .
\end{aligned}
\eeq
In the adjoint representation of SU(2), $(I_a)_{ij}=-2\,\ve_{aij}$ and $C(1)=2$, hence from
\beq
e^a I_a \= -2 \left( \begin{matrix}
0 & e^3 & -e^2 \\[4pt]
-e^3 & 0 & e^1 \\[4pt]
e^2 & -e^1 & 0
\end{matrix} \right) \quad\und\quad
\sfrac12\ve^a_{bc}\,e^b{\wedge}e^c\,I_a \= -2  \left( \begin{matrix}
0 & e^{12} & -e^{31} \\[4pt]
-e^{12} & 0 & e^{23} \\[4pt]
e^{31} & -e^{23} & 0
\end{matrix} \right)
\eeq
one finds
{\small
\beq
\begin{aligned}
I_\ch &\= -2\,\left( \begin{matrix}
0 & \cos\th & -\sin\th\cos\ph \\[4pt]
-\cos\th & 0 & \sin\th\sin\ph \\[4pt]
\sin\th\cos\ph & -\sin\th\sin\ph & 0
\end{matrix}\right)\ , \\[6pt]
I_\th &\= -2\,\sin\ch\cos\ch \left( \begin{matrix}
0 & -\sin\th &  \!\!\!\!\!\!\tan\ch\sin\th{-}\cos\th\cos\ph \\[4pt]
\sin\th & 0 & \!\!\!\!\!\!\tan\ch\cos\th{+}\cos\th\sin\ph \\[4pt]
-\tan\ch\sin\th{+}\cos\th\cos\ph & -\tan\ch\cos\th{-}\cos\th\sin\ph & 0
\end{matrix}\right) \ , \\[6pt]
I_\ph &\= -2\,\sin^2\!\ch\sin\th \left( \begin{matrix}
0 & \sin\th & \!\!\!\!\!\!\cot\ch\sin\ph{+}\cos\th\cos\ph \\[4pt]
-\sin\th & 0 &  \!\!\!\!\!\!\cot\ch\cos\ph{-}\cos\th\sin\ph \\[4pt]
-\cot\ch\sin\ph{-}\cos\th\cos\ph & -\cot\ch\cos\ph{+}\cos\th\sin\ph & 0
\end{matrix}\right) \ .
\end{aligned}
\eeq
}
{}\!\!From these expressions, it is straightforward to write down the components
of $\Acal$ and $\Fcal$ on the 3-sphere, namely $\Acal_\tau=0$ and
\beq
\begin{aligned}
\Acal_\ch &\= \sfrac12(1{+}\psi)\,I_\ch\ ,\qquad\quad\ \
\Acal_\th \= \sfrac12(1{+}\psi)\,I_\th\ ,\qquad\qquad\ \
\Acal_\ph \= \sfrac12(1{+}\psi)\,I_\ph\ ,\\[4pt]
\Fcal_{\tau\ch} &\= \sfrac12\sfrac{\diff\psi}{\diff\tau}\,I_\ch\ ,\qquad\qquad\quad
\Fcal_{\tau\th} \= \sfrac12\sfrac{\diff\psi}{\diff\tau}\,I_\th\ ,\qquad\qquad\qquad
\Fcal_{\tau\ph} \= \sfrac12\sfrac{\diff\psi}{\diff\tau}\,I_\ph\ ,\\[4pt]
\Fcal_{\ch\th} &\= \sfrac12(1{-}\psi^2)\,\sfrac1{\sin\th}\,I_\ph\ ,\quad
\Fcal_{\ch\ph} \= -\sfrac12(1{-}\psi^2)\,\sin\th\,I_\th\ ,\quad
\Fcal_{\th\ph} \= \sfrac12(1{-}\psi^2)\,\sin^2\!\ch\sin\th\,I_\ch\ .
\end{aligned}
\eeq
The corresponding electric and magnetic field components are then read off as
\beq
\begin{aligned}
E_\ch &\= \Fcal_{\tau\ch}\ ,\qquad\qquad\quad\ \
E_\th  \= \Fcal_{\tau\th}\ ,\qquad\quad\!
E_\ph  \= \Fcal_{\tau\ph}\ ,\\[4pt]
B_\ch &\= -\sfrac{1}{\sin^2\!\ch\sin\th}\,\Fcal_{\th\ph}\ ,\quad
B_\th  \= \sfrac{1}{\sin\th}\,\Fcal_{\ch\ph}\ ,\quad
B_\ph  \= -\sin\th\,\Fcal_{\ch\th}\ ,
\end{aligned}
\eeq
and we see that the geometry factors precisely cancel for the magnetic components, hence
\beq
\begin{aligned}
E_\ch &\= \sfrac12\sfrac{\diff\psi}{\diff\tau}\,I_\ch\ ,\qquad\qquad\!
E_\th  \= \sfrac12\sfrac{\diff\psi}{\diff\tau}\,I_\th\ ,\qquad\qquad\!
E_\ph  \= \sfrac12\sfrac{\diff\psi}{\diff\tau}\,I_\ph\ ,\\[4pt]
B_\ch &\= -\sfrac12(1{-}\psi^2)\,I_\ch\ ,\quad
B_\th  \= -\sfrac12(1{-}\psi^2)\,I_\th\ ,\quad
B_\ph  \= -\sfrac12(1{-}\psi^2)\,I_\ph\ .
\end{aligned}
\eeq
Inspecting the matrices we see that all components are completely regular.

To view our fields on de Sitter space, we introduce an orthonormal basis on dS$_4$,
\beq
\tilde e^0\ :=\ \diff\tilde\tau \= R\,\diff\tau \und
\tilde e^a\ :=\ R\,\cosh\tau\,e^a\ ,
\eeq
and expand
\beq
\Acal \= \Act_a\,\tilde{e}^a \qquad\textrm{and}\qquad
\Fcal \= \Fct_{0a}\,\tilde{e}^0\we \tilde{e}^a + \sfrac12 \Fct_{bc}\,\tilde{e}^b\we \tilde{e}^c
\eeq
so that
\beq
\Acal_a = R\cosh\tau\,\Act_a\ ,\qquad
\Fcal_{bc} = R^2\cosh^2\!\tau\,\Fct_{bc}\ ,\qquad
\Fcal_{\tau a} = R^2\cosh\tau\,\Fct_{0a}\ .
\eeq
Therefore, the electric and magnetic field components in closed-slicing coordinates are
\beq
\begin{aligned}
\Et_\ch &\= \sfrac12\,\sfrac{1}{R^2\cosh\tau}\,\sfrac{\diff\psi}{\diff\tau}\,I_\ch\ ,\quad\quad\!\!
\Et_\th  \= \sfrac12\,\sfrac{1}{R^2\cosh\tau}\,\sfrac{\diff\psi}{\diff\tau}\,I_\th\ ,\quad\quad\!\!
\Et_\ph  \= \sfrac12\,\sfrac{1}{R^2\cosh\tau}\,\sfrac{\diff\psi}{\diff\tau}\,I_\ph\ ,\\[4pt]
\Bt_\ch &\= -\sfrac12\,\sfrac{1{-}\psi^2}{R^2\cosh^2\!\tau}\,I_\ch\ ,\quad\quad
\Bt_\th  \= -\sfrac12\,\sfrac{1{-}\psi^2}{R^2\cosh^2\!\tau}\,I_\th\ ,\quad\quad
\Bt_\ph  \= -\sfrac12\,\sfrac{1{-}\psi^2}{R^2\cosh^2\!\tau}\,I_\ph\ .
\end{aligned}
\eeq
For our orthonormal frame, the electric and magnetic energy densities become
\beq
\begin{aligned}
\tilde\rho_e &\= -\sfrac14\,\tr\Et_a\Et_a \=
-\sfrac14\,\tr\bigl\{
E_\ch^2+\sfrac1{\sin^2\!\ch}E_\th^2+\sfrac1{\sin^2\!\th\sin^2\!\th}E_\ph^2\bigr\}
\= \sfrac{3\,C(j)}{4\,R^4\cosh^2\!\tau}\bigl(\sfrac{\diff\psi}{\diff\tau}\bigr)^2\ ,\\[4pt]
\tilde\rho_m &\= -\sfrac14\,\tr\Bt_a\Bt_a \=
-\sfrac14\,\tr\bigl\{
B_\ch^2+\sfrac1{\sin^2\!\ch}B_\th^2+\sfrac1{\sin^2\!\th\sin^2\!\th}B_\ph^2\bigr\}
\= \sfrac{3\,C(j)}{4\,R^4\cosh^4\!\tau}\bigl(1{-}\psi^2\bigr)^2\ .
\end{aligned}
\eeq
The de Sitter energy of the Yang--Mills configuration then turns out to be
($\dot\psi=\cosh\tau\,\sfrac{\diff\psi}{\diff\tau}$)
\beq
{\cal E}_{\tilde\tau} \=
\int_{S^3_R}\!\tilde e^1{\wedge}\tilde e^2{\wedge}\tilde e^3\;(\tilde\rho_e+\tilde\rho_m)
\= \sfrac34\,C(j)\,\textrm{vol}(S^3_1)\,\sfrac{\dot\psi^2+(1{-}\psi^2)^2}{R\,\cosh\tau}
\= \sfrac{3}{2}\pi^2\,C(j)\,\sfrac{1}{R\,\cosh\tau}\ .
\eeq

\subsection{Yang--Mills configuration in Hopf coordinates}

The Hopf coordinates on $S^3$ described in (\ref{2.8})-(\ref{A.5}) allow for a somewhat simpler form of 
the matrices $I_*$ in the decomposition (\ref{Imatrices}) by exploiting~(\ref{2.9}).
We abbreviate $\tilde\chi := 2\chi{-}\phi$.

In the fundamental SU(2) representation, the potential and the field strength are given by
\beq
\Acal \=
-\sfrac{\im}{4}\, (1+\psi)\,\biggl( \begin{matrix}
\diff\tilde\ch+\cos\th\,\diff\ph & \ep^{\im\tilde\ch}(-\im\diff\th-\sin\th\,\diff\ph) \\[6pt]
\ep^{-\im\tilde\ch}(\im\diff\th-\sin\th\,\diff\ph) & -\diff\tilde\ch-\cos\th\,\diff\ph
\end{matrix} \biggr) \quad\und
\eeq
\beq
\Fcal \= -\sfrac{\im}{4}\dot\psi\,\diff\tau\wedge \biggl( \begin{matrix}
\diff\tilde\ch+\cos\th\,\diff\ph & \ep^{\im\tilde\ch}(-\im\diff\th-\sin\th\,\diff\ph)
\\[6pt]
\ep^{-\im\tilde\ch}(\im\diff\th-\sin\th\,\diff\ph) & -\diff\tilde\ch-\cos\th\,\diff\ph
\end{matrix} \biggr) \ +
\eeq
\beq\nonumber
+\ \sfrac{\im}{4}(1{-}\psi^2)\left( \begin{matrix}
\sin\th\,\diff\th\we\diff\ph
& \!\!\!\!
\ep^{\im\tilde\ch}(\diff\th\we\diff\tilde\ch{-}\im\sin\th\,\diff\ph\we\diff\tilde\ch{+}
\cos\th\,\diff\th\we\diff\ph) \\[6pt] 
\ep^{{-}\im\tilde\ch}(\diff\th\we\diff\tilde\ch{+}\im\sin\th\,\diff\ph\we\diff\tilde\ch{+}
\cos\th\,\diff\th\we\diff\ph)
& \!\!\!\!
{-}\sin\th\,\diff\th\we\diff\ph
\end{matrix} \right)\ .
\eeq
In the adjoint representation, we arrive at
\beq
\Acal \= \sfrac12(1{+}\psi)\!\left(\! \begin{matrix}
0 & {-}\diff\tilde\ch{-}\cos\th\,\diff\ph & \cos\tilde\ch\,\diff\th{+}
\sin\tilde\ch\sin\th\,\diff\ph \\[4pt]
\diff\tilde\ch{+}\cos\th\,\diff\ph & 0 & -\sin\tilde\ch\,\diff\th+\cos\tilde\ch\sin\th\,\diff\ph \\[4pt]
{-}\cos\tilde\ch\,\diff\th{-}\sin\tilde\ch\sin\th\,\diff\ph &
  \sin\tilde\ch\,\diff\th{-}\cos\tilde\ch\sin\th\,\diff\ph & 0
\end{matrix} \!\!\right)\ .
\eeq
The expression for the field strength is straightforward but too lengthy to write down here.

\subsection{Yang--Mills configuration in static slicing}

With a common 2-sphere parametrization, the relation between the closed and the static slice
is given by just two relations, e.g.
\bea
\sinh\tau &=& \sqrt{1{-}\rho^2}\,\sinh t \ = \ \cos\al\,\sinh t\ ,\\[4pt]
\sin\ch\,\cosh\tau &=& \qquad \rho \qquad\qquad\! = \ \sin\al \ ,
\eea
{}from which one derives other relations, such as
\beq
\cos\ch\,\cosh\tau \= \cos\al\,\cosh t\ .
\eeq
A frequently used combination is
\beq
\Delta^2\ :=\ \cosh^2\!\tau \= 1 + \cos^2\!\al\,\sinh^2\!t
\= \cosh^2\!t-\sin^2\!\al\,\sinh^2\!t
\= \sin^2\!\al+\cos^2\!\al\,\cosh^2\!t\ .
\eeq
We may express the closed-slicing coordinates in terms of the static-slicing ones,
\beq
\tau \= \textrm{arsinh}(\cos\al\,\sinh t) \und
\ch \= \arcsin(\sin\al\;\Delta^{-1})\ ,
\eeq
with $\Delta=\Delta(\al,t)$.
{}From this it is straightforward to derive
\beq
\begin{aligned}
\frac{\pa\tau}{\pa t} &\= \frac{\cos\al\,\cosh t}{\Delta}\quad ,\qquad
\frac{\pa\tau}{\pa\al}\= -\frac{\sin\al\,\sinh t}{\Delta}\quad,\\[4pt]
\frac{\pa\ch}{\pa t}  &\= -\frac{\sin\al\,\cos\al\,\sinh t}{\Delta^2}\quad,\qquad
\frac{\pa\ch}{\pa\al} \= \frac{\cosh t}{\Delta^2}\quad,
\end{aligned}
\eeq
which provides the Jacobian for the change of `closed' to `static' variables.

In order to evaluate the components of the gauge potential and the field strength
in the static coordinates~$(x^I)$,
we transform them from the closed coordinates~$(x^i)$ according to
\beq
\Acal_I \= \sfrac{\pa x^i}{\pa x^I}\,\Acal_i \und
\Fcal_{IJ} \= \sfrac{\pa x^i}{\pa x^I}\sfrac{\pa x^j}{\pa x^J}\,\Fcal_{ij}
\eeq
and re-express the arguments $x^i=x^i(x^I)$, e.g.
\beq
\sin\ch\ \Rightarrow\ \Delta^{-1} \sin\al \und
\cos\ch\ \Rightarrow\ \Delta^{-1} \cos\al\,\cosh t\ .
\eeq
Since the $S^2$ coordinates $\th$ and~$\ph$ are common to both systems
and we employ the $\Acal_\tau=0$ gauge, we remain with
\beq
\Acal_t \= \sfrac{\pa\ch}{\pa t}\,\Acal_\ch \und
\Acal_\rho \= \sfrac1{\cos\al}\,\sfrac{\pa\ch}{\pa\al}\,\Acal_\ch
\eeq
and, since the determinant of the Jacobian equals $\Delta^{-1}$,
\beq
\begin{aligned}
\Fcal_{t\rho} &\= \sfrac{1}{\Delta}\,\Fcal_{\tau\ch}\ ,\quad
\Fcal_{t\th}   \= \sfrac{\pa\tau}{\pa t}\,\Fcal_{\tau\th}+\sfrac{\pa\ch}{\pa t}\,\Fcal_{\ch\th}\ ,\quad
\Fcal_{t\ph}   \= \sfrac{\pa\tau}{\pa t}\,\Fcal_{\tau\ph}+\sfrac{\pa\ch}{\pa t}\,\Fcal_{\ch\ph}\ ,\\[4pt]
\Fcal_{\rho\th} &\= \sfrac1{\cos\al}
\bigl(\sfrac{\pa\tau}{\pa\al}\,\Fcal_{\tau\th}+\sfrac{\pa\ch}{\pa\al}\,\Fcal_{\ch\th}\bigr)\ ,\quad
\Fcal_{\rho\ph} \= \sfrac1{\cos\al}
\bigl(\sfrac{\pa\tau}{\pa\al}\,\Fcal_{\tau\ph}+\sfrac{\pa\ch}{\pa\al}\,\Fcal_{\ch\ph}\bigr)\ .
\end{aligned}
\eeq
In these coordinates electric and magnetic field components are defined as
\beq
\begin{aligned}
E_\rho &\= \Fcal_{t\rho}\ ,\qquad\qquad\ \
E_\th   \= \Fcal_{t\th}\ ,\qquad\quad\;
E_\ph   \= \Fcal_{t\ph}\ ,\\[4pt]
B_\rho &\=  -\sfrac{1}{\rho^2\sin\th}\,\Fcal_{\th\ph}\ ,\quad
B_\th  \= \sfrac{\cos^2\!\al}{\sin\th}\,\Fcal_{\rho\ph}\ ,\quad
B_\ph  \= -\cos\al\sin\th\,\Fcal_{\rho\th}\ .
\end{aligned}
\eeq
Passing to the dimensional (tilded) coordinates multiplies these relations
with a factor of 
\beq
R^{-2}\cosh^{-2}\!\tau \= R^{-2}\Delta^{-2}\ .
\eeq
The radial components expressed in terms of the static coordinates take a reasonably simple form,
\beq
\widetilde{E}_r \= \sfrac{1}{R\,\Delta}\,\Fct_{\tau\ch}\=
\sfrac12\,\sfrac{1}{R^3\Delta^3}\,{\dot\psi}^2 I_\ch  \quad\und\quad
\widetilde{B}_r \= -\sfrac{1}{R\,\rho^2\sin\th}\,\Fct_{\th\ph}\=
-\sfrac12\,\sfrac{1}{R^3\Delta^4}(1{-}\psi^2) I_\ch \ .
\eeq
It will be interesting to physically interpret the static field components,
in particular for the limits $\al\to0$ (`center' of the configuration)
and $\al\to1$ (cosmological horizon).

\vspace{12mm}


\begin{thebibliography}{99}

\bibitem{Dirac}
P.A.M.~Dirac,
``Quantized singularities in the electromagnetic field,''\\
Proc. Roy. Soc. Lond.  A {\bf 133} (1931) 60;\\
G.~'t Hooft,
``Magnetic monopoles in unified gauge theories,''\\
Nucl. Phys.  B {\bf 79} (1974) 276;\\
A.M.~Polyakov,
``Particle spectrum in quantum field theory,''\\
JETP Lett.  {\bf 20} (1974) 194.

\bibitem{JT}
A.~Jaffe and C.~Taubes, 
{\it Vortices and monopoles}, 
Birkh\"auser, Boston, 1980.

\bibitem{Raj}
R.~Rajaraman,
{\it Solitons and instantons},
North-Holland, Amsterdam, 1984.

\bibitem{MS}
N.~Manton and P.~Sutcliffe,
{\it Topological solitons},\\
Cambridge University Press, Cambridge, 2004.

\bibitem{Wein}
E.J.~Weinberg,
{\it Classical solutions in quantum field theory},\\
Cambridge University Press, Cambridge, 2015.

\bibitem{KK}
B.~Kleihaus, J.~Kunz and F.~Navarro-Lerida,
``Rotating black holes with non-Abelian hair,''
Class.\ Quant.\ Grav.\  {\bf 33} (2016) 234002
[arXiv:1609.07357 [hep-th]].

\bibitem{IvLePo}
T.A.~Ivanova, O.~Lechtenfeld and A.D.~Popov,
``Solutions to Yang-Mills equations on four-dimensional de Sitter space,''
Phys.\ Rev.\ Lett.\ {\bf 119} (2017) 061601 
[arXiv:1704.07456 [hep-th]].

\bibitem{HE}
S.W.~Hawking and G.F.R.~Ellis,
{\it The large scale structure of space-time,}\\
Cambridge University Press, Cambridge, 1975.

\bibitem{Don}
S.K.~Donaldson,
``Boundary value problems for Yang-Mills fields,''\\
J.\ Geom.\ Phys.\  {\bf 8} (1992) 89.

\bibitem{IL}
T.A.~Ivanova and O.~Lechtenfeld,
``Yang-Mills instantons and dyons on group manifolds,''\\
Phys.\ Lett.\ B {\bf 670} (2008) 91
[arXiv:.0806.0394 [hep-th]].

\bibitem{ILPR}
T.A.~Ivanova, O.~Lechtenfeld, A.D.~Popov and T.~Rahn,
``Instantons and Yang-Mills flows on coset spaces,''
Lett.\ Math.\ Phys.\  {\bf 89} (2009) 231
[arXiv:0904.0654 [hep-th]].

\bibitem{BILL}
I.~Bauer, T.A.~Ivanova, O.~Lechtenfeld and F.~Lubbe,
``Yang-Mills instantons and dyons on homogeneous G$_2$-manifolds,''
JHEP {\bf 10} (2010) 044
[arXiv:1006.2388 [hep-th]].

\bibitem{ILPS}
T.A.~Ivanova, O.~Lechtenfeld, A.D.~Popov and R.J.~Szabo,\\
``Orbifold instantons, moment maps and Yang-Mills theory with sources,''\\
Phys.\ Rev.\ D {\bf 88} (2013) 105026
[arXiv:1310.3028 [hep-th]].

\bibitem{LPS}
O.~Lechtenfeld, A.D.~Popov and R.J.~Szabo,
``Sasakian quiver gauge theories and instantons on Calabi-Yau cones,''
Adv.\ Theor.\ Math.\ Phys.\  {\bf 20} (2016) 821
[arXiv:1412.4409 [hep-th]].

\bibitem{PH}
J.~Podolsky and O.~Hruska,
``Yet another family of diagonal metrics for de Sitter and anti–de Sitter spacetimes,''
Phys.\ Rev.\ D {\bf 95} (2017) 124052
[arXiv:1703.01367 [gr-qc]].

\bibitem{MaS}
N.S.~Manton and T.M.~Samols,
``Sphalerons on a circle,''
Phys.\ Lett.\ B {\bf 207} (1988) 179.

\bibitem{LMKT}
J.Q.~Liang, H.J.W.~Muller-Kirsten and D.H.~Tchrakian,\\
``Solitons, bounces and sphalerons on a circle,''
Phys.\ Lett.\ B {\bf 282} (1992) 105.

\bibitem{ale}
D.V.~Alekseevsky, V.~Cort\'es, A.~Galaev and T.~Leistner,
``Cones over pseudo-Riemannian manifolds and their holonomy''
J. Reine Angew. Math. {\bf 635} (2009) 23  [arXiv:0707.3063 [math.DG]].

\bibitem{ABT}
O.~Aharony, M.~Berkooz, D.~Tong and S.~Yankielowicz,\\
``Confinement in anti-de Sitter space,''
JHEP {\bf 02} (2013) 076
[arXiv:1210.5195 [hep-th]].

\bibitem{CW}
C.G.~Callan, Jr. and F.~Wilczek,
``Infrared behavior at negative curvature,''\\
Nucl.\ Phys.\ B {\bf 340} (1990) 366.

\bibitem{AFF}
V.~de Alfaro, S.~Fubini and G.~Furlan,\\
``A new classical solution of the Yang-Mills field equations,''
Phys.\ Lett.\ B {\bf 65} (1976) 163.

\bibitem{Popov}
A.D.~Popov,
``Hermitian-Yang-Mills equations and pseudo-holomorphic bundles on nearly K\"ahler and nearly
Calabi-Yau twistor 6-manifolds,''\\
Nucl.\ Phys.\ B {\bf 828} (2010) 594
[arXiv:0907.0106 [hep-th]].


\end{thebibliography}
\end{document}